\documentclass[preprint2]{aastex6}
\usepackage[fleqn]{amsmath}
\usepackage{amssymb}

\slugcomment{Not to appear in Nonlearned J., 45.}

\shorttitle{The High cadence Transient Survey (HiTS)} \shortauthors{F\"orster et al.}

\begin{document}


\title{The High cadence Transient Survey (HiTS)-- I. Survey design and
  supernova shock breakout constraints.}


\author{F.~F\"orster\altaffilmark{1, 2},
  J.C.~Maureira\altaffilmark{1},
  J.~San Mart\'in\altaffilmark{1},
  M.~Hamuy\altaffilmark{3, 2},
  J.~Mart\'inez\altaffilmark{3, 2},
  P.~Huijse\altaffilmark{2, 4},
  G.~Cabrera\altaffilmark{2, 1, 5},
  L.~Galbany\altaffilmark{2, 3},
  Th.~de Jaeger\altaffilmark{2, 3},
  S.~Gonz\'alez--Gait\'an\altaffilmark{1, 2},
  J.P.~Anderson\altaffilmark{6},
  H.~Kunkarayakti\altaffilmark{2, 3},
  G.~Pignata\altaffilmark{7, 2},
  F.~Bufano\altaffilmark{8, 2},
  J.~Litt\'in\altaffilmark{1},
  F.~Olivares\altaffilmark{7},
  G.~Medina\altaffilmark{3},
  R.C.~Smith\altaffilmark{5},
  A.K. Vivas\altaffilmark{5},
  P.A. Est\'evez\altaffilmark{4, 2},
  R.~Mu\~noz\altaffilmark{3},
  E.~Vera\altaffilmark{1}
}
\email{francisco.forster@gmail.com}


\altaffiltext{1}{Center for Mathematical Modelling, University of Chile, Beaucheff 851, Santiago, Chile.}
\altaffiltext{2}{Millennium Institute of Astrophysics, Chile.}
\altaffiltext{3}{Departamento de Astronom\'ia, Universidad de Chile, Camino el Observatorio 1515, Santiago, Chile.}
\altaffiltext{4}{Departamento de Ingenier\'ia El\'ectrica, Universidad de Chile, Casilla 412-3, Santiago, Chile.}
\altaffiltext{5}{Cerro Tololo Inter-American Observatory, National Optical Astronomy
Observatory, Casilla 603, La Serena, Chile}
\altaffiltext{6}{European Southern Observatory, Alonso de Córdova 3107, Vitacura, Santiago, Chile}
\altaffiltext{7}{Departamento de Ciencias Fisicas, Universidad Andres Bello, Avda. Republica 252, Santiago, Chile}
\altaffiltext{8}{INAF - Osservatorio Astrofisico di Catania, Via Santa Sofia, 78, Catania, Italy}


\begin{abstract}
We present the first results of the High cadence Transient Survey
(HiTS), a survey whose objective is to detect and follow up optical
transients with characteristic timescales from hours to days,
especially the earliest hours of supernova (SN) explosions. HiTS uses
the Dark Energy Camera (DECam) and a custom made pipeline for image
subtraction, candidate filtering and candidate visualization, which
runs in real--time to be able to react rapidly to the new transients.
We discuss the survey design, the technical challenges associated with
the real--time analysis of these large volumes of data and our first
results.  In our 2013, 2014 and 2015 campaigns we have detected more
than 120 young SN candidates, but we did not find a clear signature
from the short--lived SN shock breakouts (SBOs) originating after the
core collapse of red supergiant stars, which was the initial science
aim of this survey. Using the empirical distribution of
limiting--magnitudes from our observational campaigns we measured the
expected recovery fraction of randomly injected SN light curves which
included SBO optical peaks produced with models from
\citet{2011ApJS..193...20T} and \citet{2010ApJ...725..904N}. From this
analysis we cannot rule out the models from
\citet{2011ApJS..193...20T} under any reasonable distributions of
progenitor masses, but we can marginally rule out the brighter and
longer--lived SBO models from \citet{2010ApJ...725..904N} under our
best--guess distribution of progenitor masses. Finally, we highlight
the implications of this work for future massive datasets produced by
astronomical observatories such as LSST.

\end{abstract}



\section{INTRODUCTION}

The advent of a new generation of large field--of--view astronomical
optical CCD cameras already operational (e.g. iPTF,
\citealt{2009PASP..121.1395L}; SkyMapper,
\citealt{2007ASPC..364..177K}; Pan--STARRS,
\citealt{2004AN....325..636H}; OmegaCam,
\citealt{2002Msngr.110...15K}; DECam, \citealt{2015AJ....150..150F};
Hyper Suprime--Cam, \citealt{2010SPIE.7740E..2IF}; KMTNET,
\citealt{2011SPIE.8151E..1BK}) and planned (e.g. ZTF,
\verb+http://www.ptf.caltech.edu/ztf+; and LSST,
\citealt{2009arXiv0912.0201L}) is revolutionizing our understanding of
the Universe because of their surveying capabilities.  Thanks to these
instruments, large regions in the sky are being mapped up to very
large distances, and also rare, short--lived optical transients are
being found as these large regions of the Universe are observed with a
high cadence. The latter presents not only new opportunities for the
study of astrophysical phenomena, but also new challenges from the
point of view of the data analysis. Large volumes of data will have to
be processed in real--time in order to trigger follow up observations
that would help disentangle the physical nature of the transients
detected \citep[see e.g.][]{2014Natur.509..471G}.

The High cadence Transient Survey (HiTS) is a discovery survey that
takes advantage of the large \emph{etendue}, the product of collecting
area and field--of--view, of the Dark Energy Camera (DECam) mounted on
the 4 m Blanco telescope at the Cerro Tololo Interamerican Observatory
(CTIO), the fast connectivity available between CTIO and the Center
for Mathematical Modelling (CMM~@~U.~Chile), and the computing
capabilities of the Chilean National Laboratory for High Performance
Computing (NLHPC) that allows us to observe and analyze high cadence
DECam data in real--time. Because DECam is the largest \emph{etendue}
project in the southern hemisphere until the arrival of the full LSST
project, HiTS can be considered a precursor project for some of the
challenges regarding the fast analysis of large data volumes, the high
cadence observations of deep drilling fields and, depending on the
cadence, the exploration of the hour--timescale transient population.

HiTS aims to explore the population of transient or periodic objects
with characteristic timescales between a few hours and days
\citep[c.f.][]{2010ApJ...723L..98K} and apparent magnitudes down to
about 24.5 mag. Its main science driver was to discover the elusive
shock breakout (SBO) phase of extended red supergiant star (RSG)
progenitors undergoing core collapse \citep{1978ApJ...225L.133F,
  2008Sci...321..223S, 2008ApJ...683L.131G}, but it also focuses on
the study of young SN explosions in general. The latter includes
shock--hit companion stars in multiple progenitor systems of Type Ia
SNe explosions \citep[see e.g.][]{2000ApJS..128..615M,
  2010ApJ...708.1025K, 2011ApJ...741...20B, 2012ApJ...744L..17B,
  2014ApJ...784L..12G, 2016ApJ...820...92M, 2015Natur.521..328C}; the
shock cooling phase of core collapse supernovae (SNe), which provide
constraints on their progenitors size and circumstellar environments
\citep{2011MNRAS.415..199M, 2015MNRAS.451.2212G, 2016ApJ...819...35A};
and the early light curves of Type Ia SNe, whose diversity could be
driven by different radioactive profiles in their outermost layers
\citep{2014ApJ...784...85P}.

The structure of this manuscript is the following: in
Section~\ref{sec:sne} we will describe some of the relevant physics of
red supergiant SN explosions during their earliest observable phases;
in Section~\ref{sec:design} we will discuss how the survey was
designed and the details of our observation strategy; in
Section~\ref{sec:data} we will show how the real--time analysis of the
data was performed, including a brief description of a newly developed
image subtraction, candidate classification and visualization
pipeline; in Section~\ref{sec:results} we will discuss some of the
first results, including a detailed discussion on the
limiting--magnitude of the survey and its implications for SBO model
constraints; in Section~\ref{sec:summary} we summarize the main
results from this paper and in Section~\ref{sec:discussion} we discuss
the implications from this work. We note that in this manuscript we
only present our conclusions about the presence or absence of RSG
SBOs, leaving the discussion on other classes of transient events for
subsequent publications.

\section{Core collapse supernova} \label{sec:sne}

\subsection{Shock breakout}

The core collapse of a massive star leads to the formation of a
compact object and a shock wave that can unbind its outer layers in a
core collapse supernova (SN). The shock can traverse the entire
progenitor at very high velocities, $v_{shock} \approx 0.01 ~c$, until
it eventually emerges at the surface of the star in a shock breakout
(SBO) event. The shock's energy density will be dominated by radiation
and its main source of opacity will be Compton scattering by free
electrons \citep{1976ApJS...32..233W}. The shock thickness, $d_{\rm
  shock}$, can be estimated by equating the photon diffusion
timescale, the typical time taken by photons to random walk out of the
shock via electron scattering, and the advection timescale, the time
taken by the shock to traverse one shock thickness. This results in a
shock thickness of approximately $d_{\rm shock} \sim \frac{c
  ~l_{s}}{v_{\rm shock}}$, where $l_s$ is the electron scattering mean
free path and $v_{\rm shock}$ is the shock speed, or an optical depth
of $\tau_{\rm shock} \sim \frac{c}{v_{\rm shock}}$ \citep[ignoring
  form factors that may be significant in some cases, see
  e.g.][]{2008Sci...321..223S}.

As the shock front approaches the stellar surface it will encounter a
decreasing optical depth. When $\tau \le \tau_{\rm shock}$ the
radiation dominated shock will leak its energy out of the stellar
surface as a radiative precursor until the shock breaks out completely
and becomes rarified. The shock breakout timescale, $t_{\rm shock}$,
will be given by the time that it takes for the radiative precursor to
leak out into the stellar surface, $t_{\rm shock} \sim \frac{d_{\rm
    shock}}{v_{\rm shock}}$, but for an external observer it will also
be affected by the light crossing time, $t_{\rm light} \sim
\frac{R_{\star}}{c}$, where $R_{\star}$ is the star's radius. Shock
breakout timescales are expected to be typically of about an hour for
RSGs; several minutes for blue supergiants (BSGs); and several seconds
for stripped--envelope stars (\citealt{2013ApJ...778...81K}, c.f.
\citealt{2008Sci...321..223S}).

The shock breakout properties will depend on many structure and
composition parameters, but most strongly on the radius of the
progenitor star \citep{2004MNRAS.351..694C}. Stripped--envelope, BSG
and RSG stars have very different radii: of the order of $5-10
~R_{\odot}$, $25-50 ~R_{\odot}$ and $500-1000 ~R_{\odot}$,
respectively. Even for similar masses, the envelope structure can vary
significantly depending on whether the envelope energy transport is
radiative or convective. Because BSG stars have radiative envelopes
and RSG stars, convective envelopes, they have very different
effective polytropic indices ($n = 3$ in BSGs and $n = 1.5$ in RSGs),
which also leads to different shock acceleration
dynamics. Additionally, the presence of pre--SN winds can strongly
change the shock propagation physics and the observable properties of
the breakout event \citep{2011MNRAS.415..199M, 2011MNRAS.414.1715B,
  2012ApJ...759..108S, 2014ApJ...788..113S}.

The characteristic temperature of radiation during breakout depends
mainly on the radius of the progenitor stars, being aproximately
proportional to the inverse of the progenitor radius squared
\citep{2004MNRAS.351..694C}, but the typical energy of the photons
leaving the star will also depend on whether thermal equilibrium
between the escaping radiation deep inside the star and the
intervening gas is achieved and if so, the depth at which this occurs
\citep{2010ApJ...725..904N}. Most photons leaving the star during
shock breakout will be in X--ray and UV wavelengths, making the
detection of these events very difficult from the ground. In fact, a
few shock breakout candidates have been detected from space: three SNe
II observed in the UV at early times with \emph{GALEX}
\citep{2008Sci...321..223S, 2008ApJ...683L.131G} and a recent SN II
observed in the optical using \emph{Kepler}
\citep{2016ApJ...820...23G} have been associated with shocks breaking
out of the envelopes of their RSG progenitors. Other SNe II appear to
have shocks breaking into high density circumstellar material
\citep{2010ApJ...720L..77G, 2015ApJ...804...28G, 2015MNRAS.451.2212G,
  2016ApJ...818....3K, 2016arXiv160103261T, 2016ApJ...820...23G},
which is supported by models of red supergiant winds
\citep{2014Natur.512..282M}. A \emph{Swift} X--ray transient has been
associated with an stripped--envelope SNe Ibc, from what appears to be
a Wolf Rayet star \citep{2008Natur.453..469S} and/or its surrounding
wind \citep{2014ApJ...788L..14S}. No direct detections of shock
breakouts from BSG stars have been made, but there is indirect
evidence for a shock breakout in SN 1987A \citep{1996ApJ...464..924L}.

Although UV and X--ray detectors are better suited for shock breakout
detections because of the typical temperatures encountered at the
shock's front, \citet{2011ApJS..193...20T} suggested that optical
detectors can also be used to find these events. In fact, their models
suggest that it may be easier to detect SBOs in a systematic way with
a new generation of large field--of--view optical CCD cameras \citep[see
  e.g.][]{2014PASJ...66..114M, 2016ApJ...820...23G, SHOOTS}.

\subsection{Shock cooling, plateau and radioactive tail}

After shock breakout, the outer layers of the star will enter an
adiabatic expansion and cooling phase, the so--called shock cooling
phase. \citet{2010ApJ...725..904N} have divided the adiabatic
expansion evolution after breakout into two distinct phases: planar
and spherical expansion. During planar expansion the dominant
contribution to the luminosity comes from the breakout shell, which
would evolve adiabatically due to expansion at almost constant radius
and with a luminosity $\propto t^{-4/3}$. During spherical expansion
the radius cannot be considered constant and energy from the inner
shells also contributes to the light curve, with a slower luminosity
evolution $\propto t^{-0.17}$ to $t^{-0.35}$ for polytropic indices
$1.5 \le n \le 3$. For RSGs, the transition between planar and
spherical expansion is expected at about 14 hr after explosion.

The evolution of the near ultraviolet (NUV) and optical luminosity
during adiabatic expansion can differ significantly from that of the
total luminosity.  While the total luminosity generally declines
monotonically after shock breakout, the NUV and optical luminosity can
decline and then rise again for several days up to maximum optical
light \citep[e.g.][]{2008Sci...321..223S, 2011ApJS..193...20T}.

The adiabatic approximation will be valid preferentially in the denser
inner layers, as the outer layers become dominated by radiative
diffusion with time.  The radius where this transition occurs is known
as the diffusion wave radius, which can be estimated by equating the
star's radiation diffusion time with the time since explosion. In
those stars where radiative diffusion is negligible during this phase,
with initial radii of $R_{\star} \le 100 ~R_{\odot}$, the product of
$R_{\star} ~ E_{\rm in}^{0.91} M_{\rm ej}^{-0.40}$ can be constrained
by the luminosity evolution, where $R_{\star}$ is the stellar radius,
$E_{\rm in}$ is the explosion energy and $M_{\rm ej}$ is the mass of
the ejecta \citep{1992ApJ...394..599C}.

\begin{figure}[!ht]
 \centering
\includegraphics[scale=0.55]{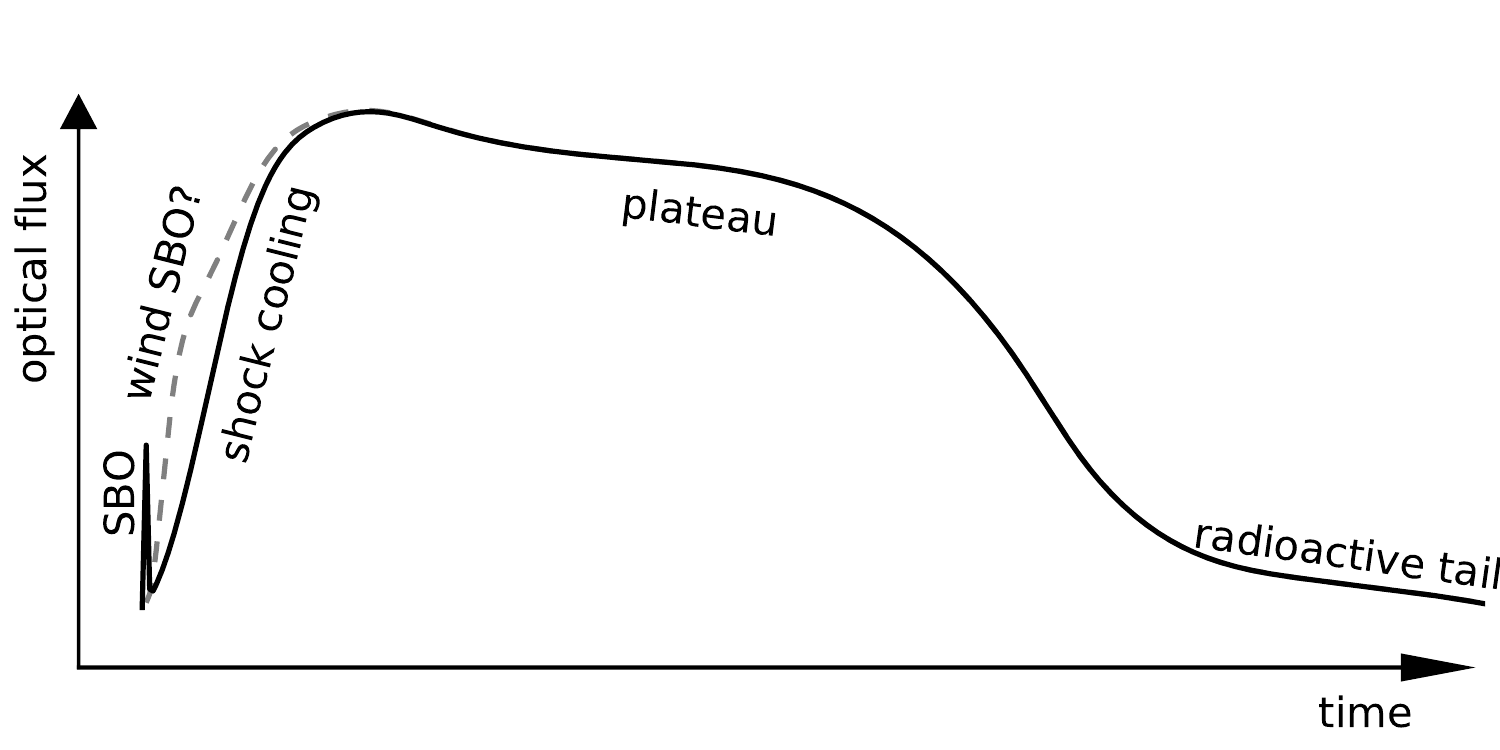} 
\caption{Schematic representation of the optical light curve of a red
  supergiant core collapse supernova in arbitrary units (c.f. Figure 1
  in \citealt{2016ApJ...820...23G}). Note that the wind SBO optical
  flux depends on the properties of the circumstellar material and may
  not always be dominant at early times. }
\label{fig:sn}
\end{figure}

As radiative diffusion becomes important the ionized He and H envelope
can recombine in a wave which sweeps over the star inwards in mass
coordinates, dominating the luminosity evolution in the plateau
phase. The time when this transition occurs depends on the envelope
mass and its structure, but it will usually start a few days after
explosion \citep{1992ApJ...394..599C}. The luminosity during this
phase can evolve relatively slowly for a few months \citep[see
  e.g.][]{2012ApJ...756L..30A, 2014ApJ...786...67A}, depending on the
explosion energy, ejected mass, initial radius and composition
\cite[see e.g.][]{1993ApJ...414..712P, 2009ApJ...703.2205K}, and will
be followed by an exponentially decaying radioactive tail phase of
evolution explained by the presence of newly synthesized $^{56}$Ni. A
schematic representation of the previous phases of evolution is shown
in Figure~\ref{fig:sn}.

\section{SURVEY DESIGN} \label{sec:design}

Designing a real--time survey to look for supernova (SN) shock
breakouts (SBOs) can be seen as an optimization problem where the
objective function is the total number of events to be detected. The
constraints are that the cadence should be similar to the typical
timescale of SN SBOs (of the order of an hour), that the time between
observations cannot be shorter than what can be processed in a steady
state with the available computational resources, and for our purposes
that the events cannot be located at too large distances in order to
facilitate the follow up with other astronomical instruments.

Unambiguous SBO detections should be ideally made in observational
triplets with a timescale comparable to the typical time--scale of red
supergiant (RSG) SBOs, each composed of a non--detection, detection
and non--detection/confirmation; as long as the position of the
candidate coincides with the position of a subsequent SN explosion to
discard other variable sources. Triplets can be more efficiently
obtained with a high cadence, full night strategy: five epochs in a
night contain three triplets with the same cadence, but three epochs
in a night contain only one triplet. However, requiring too many
epochs per night per field with a cadence of the order of an hour can
limit the area in the sky which can be efficiently observed due to
airmass constraints depending on the latitude of the observatory. The
SBOs and subsequent SNe can be both observed during the high cadence
phase if it spans for several nights, which should ideally relax into
a low--cadence follow up phase for several months to extract physical
parameters from the SN light curve plateau evolution. 

Most observational parameters are correlated and affect the objective
function of the optimization problem. The faster the cadence the
smaller the survey area. Shorter individual exposure times for a fixed
cadence result in a larger survey area, but in a shallower survey with
a larger fraction of overhead time, affecting the total volume of the
survey in a non--obvious way depending on the telescope being
used. Multiwavelength high cadence observations result in longer
overhead times and slower cadences in a given band for a fixed area.

In this Section we will focus on the optimization of the initial high
cadence phase, introducing two figures of merit used for this purpose.

\subsection{Survey instrument and cadence} \label{sec:surveycadence}

\begin{deluxetable}{l c c c c} 
  \tablecaption{Selection of large field--of--view (FOV) optical
    astronomical cameras. The collecting area, FOV, \emph{etendue}
    (product of area and FOV) and number of pixels of each
    survey telescope are given. All listed cameras are already operational,
    except for LSST and ZFT.}
  \tablehead{\colhead{Camera} & \colhead{Area} & \colhead{Field--of--view} & \colhead{\emph{Etendue}} & \colhead{Pixels} \\ &
    \colhead{[m$^2$]} & \colhead{[deg$^2$]} & \colhead{[m$^2$
        deg$^2$]} & \colhead{[Mpix]}} \startdata
  \emph{Kepler} & 0.7 & 115 & 81.5 & 94.6\footnote{Limited bandwidth requires pre--selection of pixels} \\
  HSC\footnote{Hyper Suprime Camera} & 52.8 & 1.5 & 79.2 & 870 \\
  DECam & 11.3 & 3.0 & 33.9 & 520 \\
  PanSTARRS--1\footnote{Panoramic Survey Telescope \& Rapid Response System}  & 2.5 & 7.0 & 17.5 & 1400 \\
  iPTF\footnote{Intermediate Palomar
    Transient Facility} & 1.1 & 7.8 & 8.6 & 92 \\
  SkyMapper & 1.4 & 5.7 & 8.2 & 256 \\
  KMTNet\footnote{Koren Microlensing Telescope Network}  & 2.0 & 4.0 & 8. & 340 \\
  QUEST\footnote{Quasar Equatorial Survey Team} & 0.8 & 8.3 & 6.5 & 40.3 \vspace{.2cm} \\
  LSST\footnote{Large Synoptic Survey Telescope} & 35.7 & 9.6 & 344.2 & 3200 \\
  ZTF\footnote{Zwicky Telescope Facility} & 1.1 & 47 & 51.7 & 576 
  \enddata
\end{deluxetable}

The most relevant variable characterizing the discovery potential of a
survey telescope is the \emph{etendue} -- the product of the
telescope's collecting area and the camera's
field--of--view. Similarly, the data analysis challenges of a
real--time survey are best characterized by the data acquisition rate,
given by the ratio between the camera's image size and the typical
exposure time plus overheads. A summary of some large field--of--view
cameras mounted in large collecting area telescopes is shown in
Table~1. The large \emph{etendue} of DECam is what
enabled HiTS, but its large pixel number combined with the required
real--time analysis became its main technical challenges.

\begin{figure*}[ht!]
\hbox{
  \includegraphics[scale=0.42]{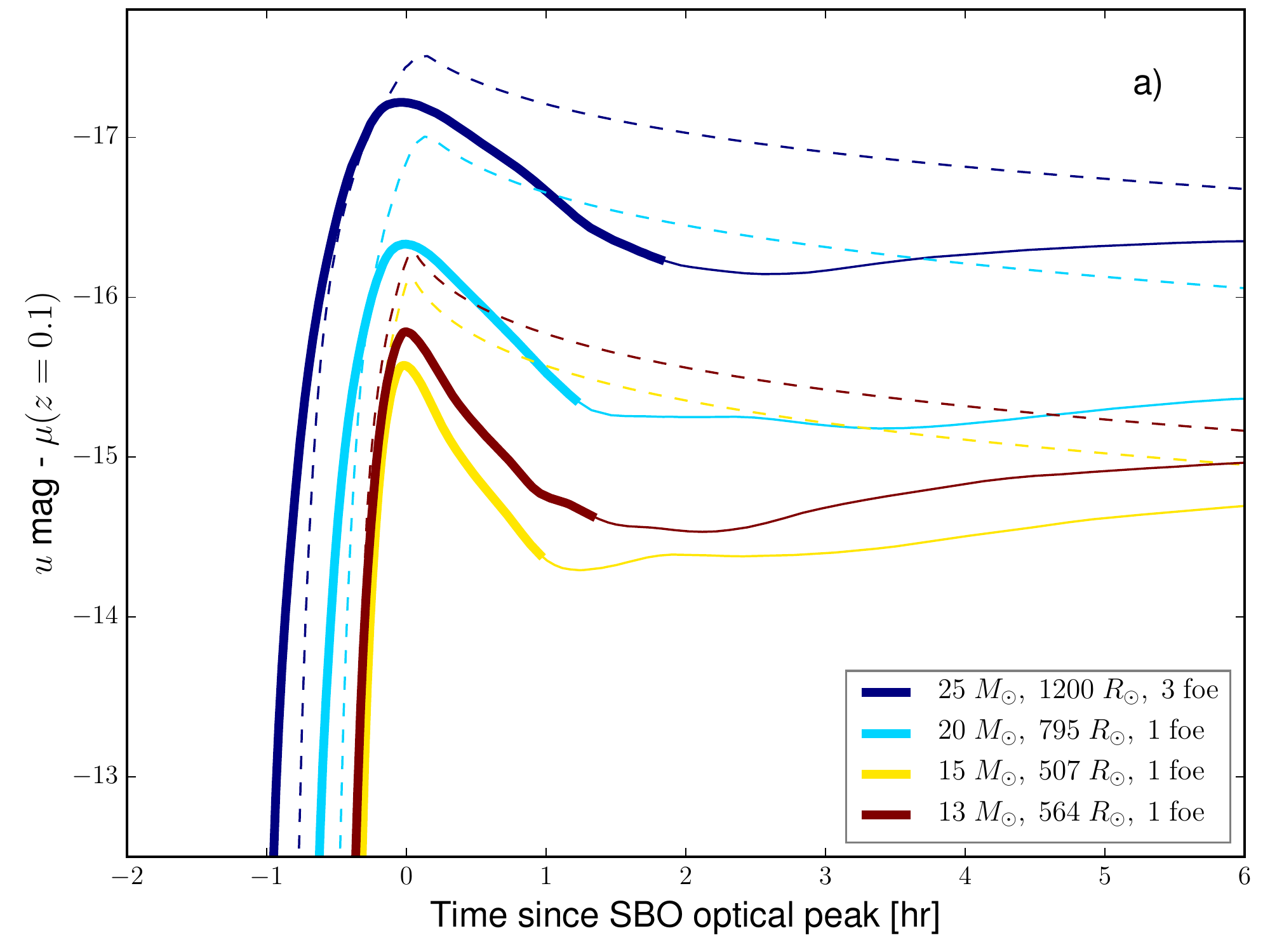}
  \includegraphics[scale=0.42]{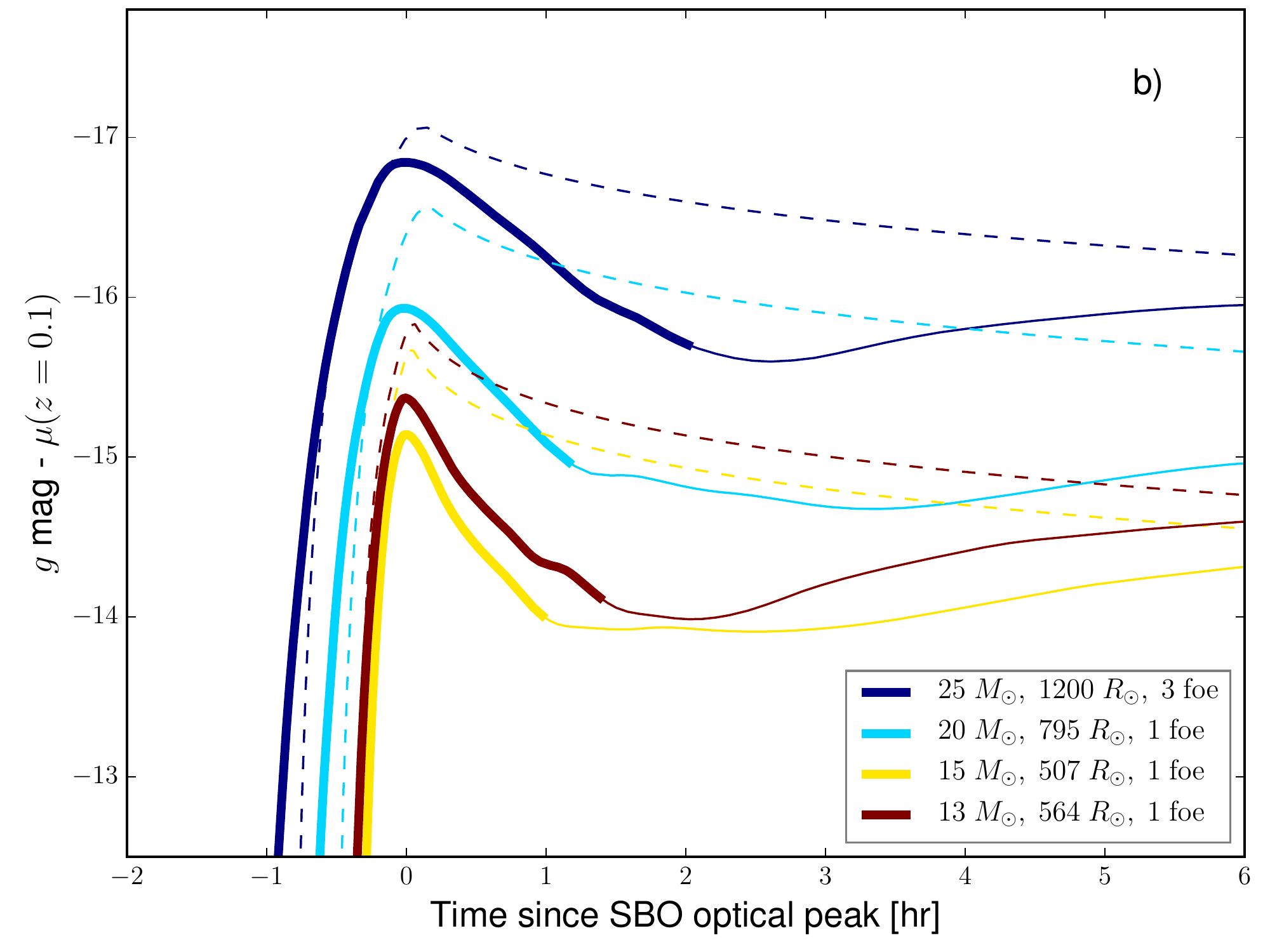}
}
\caption{Evolution of the difference between the apparent magnitude
  and the distance modulus, i.e. an \emph{effective} absolute
  magnitude, in $u$ band {\bf (a)} and $g$ band {\bf (b)}, as seen at
  redshift 0.1, from shock breakout (SBO) optical peak maxima. Based
  on models from \citealt{2011ApJS..193...20T} (continuous lines) and
  \citealt{2010ApJ...725..904N} (dashed lines) and the DECam
  filters. The legend shows the zero age main
  sequence masses ($M_{\rm ZAMS}$), the radii just before explosion
  and the explosion energy. See text for more details.}
\label{fig:SBO_LCs}
\end{figure*}

\begin{figure*}[ht!]
\hbox{
  \includegraphics[scale=0.42]{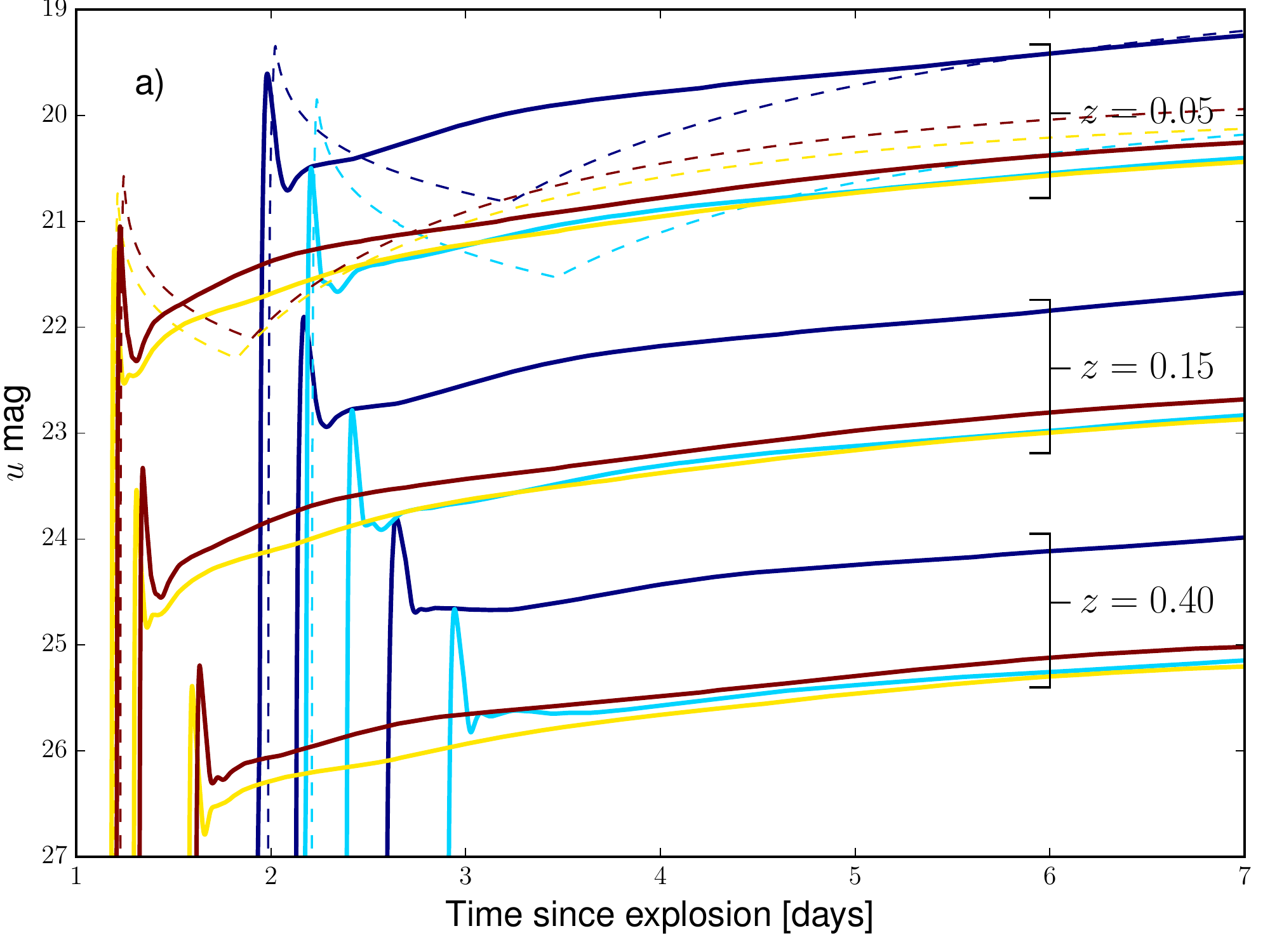} 
  \includegraphics[scale=0.42]{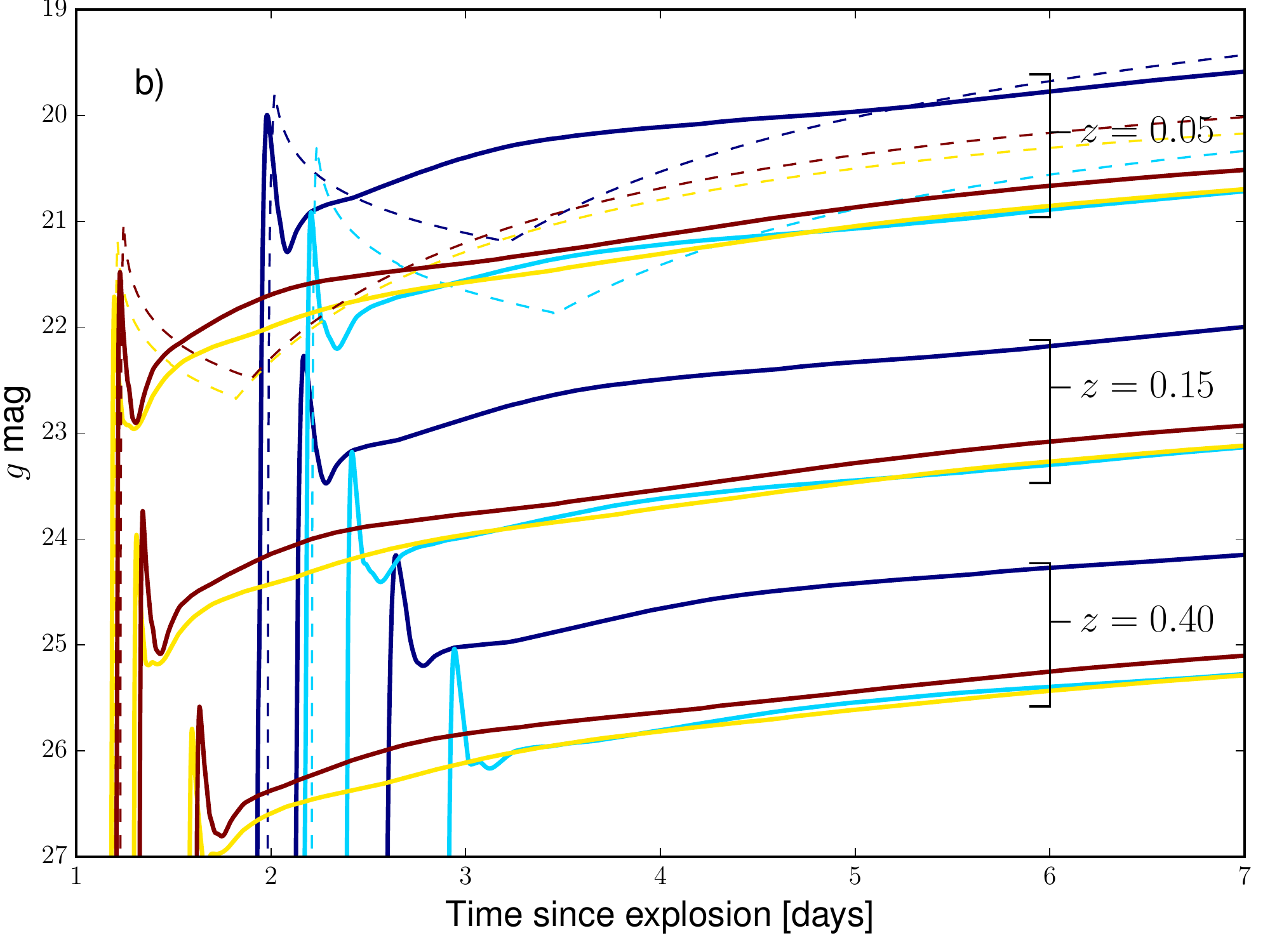} 
}
\caption{Apparent magnitude evolution in the $u$ band {\bf (a)} and
  $g$ band {\bf (b)} from explosion time as seen from different
  redshifts, based on different models from
  \citealt{2011ApJS..193...20T} (continuous lines),
  \citealt{2010ApJ...725..904N} (dashed lines), the DECam filters and
  a standard $\Lambda$--CDM cosmology. The explosion times are taken
  from the models of \citet{2011ApJS..193...20T}. The models from
  \citealt{2010ApJ...725..904N} are aligned to match the position of
  the SBO optical peak maxima and are only shown at redshift 0.05. The
  color coding is the same as in Figure~\ref{fig:SBO_LCs}.}
\label{fig:day1_LCs}
\end{figure*}

As mentioned before, the most important science driver for HiTS was
the detection of shock breakout (SBO) events in the optical in a
systematic fashion. In an initial stage the strategy consisted in
detecting SBO events in single band observations, giving preference to
rapid monochromatic sampling over multiwavelength characterization and
attempting rapid follow up of any of the detected events using our
real--time analysis and Target of Opportunity (ToO) follow up
capabilities. If the SBO detection is not achieved in real--time, the
position of the associated rising SNe could be used to look for the
SBO event in a post--processing, forced--photometry analysis (in which
we fix both the position and shape of the point spread function,
PSF). Then, if the number of detected events was found to be
significant, we would attempt the detection and spectral index
characterization in a second phase of multiwavelength, high cadence,
real--time analysis.

SBO characteristic timescales range from seconds to a few hours
\citep{2013ApJ...778...81K}, but we first focused on the longest
timescales among SBO events, i.e.  those originating from RSG stars
and with characteristic timescales of an hour or more, in order not to
compromise the survey effective volume. For this family of explosions
we use the RSG explosion models from \citet{2011ApJS..193...20T} and
\citet{2010ApJ...725..904N}, shown in Figures~\ref{fig:SBO_LCs} and
\ref{fig:day1_LCs} for the same progenitor parameters, with zero age
main sequence masses ($M_{\rm ZAMS}$) of 13, 15, 20 and 25 $M_{\odot}$
and explosion energies of 1 foe\footnote{1 foe = $10^{51}$ erg} for
the 13, 15 and 20 $M_\odot$ models and 3 foe for the 25 $M_\odot$
model, in order to mimic the mass--energy relation found by
\citet{2003ApJ...582..905H} for observed SNe II and with some
theoretical support from \citet{2016MNRAS.460..742M} for core collapse
SNe in general. To our knowledge, the models from
\citet{2011ApJS..193...20T} are the only set of optical SBO models
available in the literature which use exact solutions for realistic
progenitor models and with a range of progenitor properties. We also
use the analytic solutions from \citet{2010ApJ...725..904N} with the
same parameters for comparison throughout the analysis.

We consider cadences between 1.3 and 2 hours (4, 5 or 6 epochs per
field per night) in what follows, which can be considered to be on the
long side of the SBO timescale distribution, but that would allow us
to put constraints on the number of RSG SBO events that could be
detected and characterized by future optical surveys with sub--hour
cadences. For comparison, the Kepler ExtraGalactic survey
\citep{2015Natur.521..332O} has a cadence of 30 minutes and the Dark
Energy Survey time supernova survey \citep{2016MNRAS.460.1270D} has a
typical cadence of about a week.

We focused on the detection of the initial peak seen in
Figure~\ref{fig:SBO_LCs}, which we define as any detection with a
signal to noise ratio of at least 5 before 90\% of the magnitude
change between the optical peak maximum and subsequent minimum has
occurred for the \citet{2011ApJS..193...20T} models (thick lines in
Figure~\ref{fig:SBO_LCs}), or before the beginning of the spherical
phase of evolution for the \citet{2010ApJ...725..904N} models (break
in the dashed--lines in Figure~\ref{fig:day1_LCs}). To confirm that
these restrictions would lead to recognizable SBOs we simulated 2400
light curves for each family of models with different extinctions,
redshifts and progenitor properties using the empirical cadence and
limiting magnitudes of the survey and found that we are typically 97\%
and 95\% efficient in visually recognizing a SBO detection with the
previous restrictions under the models of \citet{2011ApJS..193...20T}
and \citet{2010ApJ...725..904N}, respectively, with a purity of 99\%
for both families of models.

\subsection{Targeted or blind survey strategy}

We considered whether to select the survey fields by their intrinsic
properties or not, i.e. to perform a targeted or blind survey,
respectively. Here targeted refers to targeting clusters of galaxies
instead of individual galaxies, given DECam's field of view. To choose
the best strategy we compared the outcomes of simulations of SBO
events using the volumetric SN rate from \citet{2015ApJ...813...93S}
and the cluster SN rate from \citet{2012ApJ...753...68G} with typical
galaxy cluster density profiles from \citet{2006MNRAS.368..518V}.  We
found that targeting clusters of galaxies with DECam we could obtain
an estimated 10\% increase in the number of events compared to an
untargeted survey.  Given that there is a relatively small number of
known clusters at relatively low redshifts
\citep{2013MNRAS.429.3272C}, of which only a small fraction remains
visible for the entire night, we estimate that the time spent slewing
the telescope from target to target would be more than 10\% of the
photon collecting time of a blind survey. Therefore, we decided to
perform a blind search, targeting cluster or supercluster fields only
when they were adjacent to some of our blind fields, which would also
allow us to obtain volumetric SN rates with fewer selection biases.

\subsection{Field configuration} 

Having selected a blind field strategy we find a reference right
ascension (RA) that makes our targets achieve the lowest possible
airmasses during all epochs, which is determined by the time of the
year when the observations were initially allocated. Then we find a
reference declination (DEC) that minimizes the average total
extinction. We use an approximately rectangular grid of closely
separated fields in the sky around this position, always switching to
adjacent fields to minimize slew times \footnote{see animations in
  https://github.com/fforster/HiTS-public}, and balancing the exposure
times with the rectangle dimensions to achieve an approximately
constant airmass during each epoch of our observations. For example,
we can observe along an approximately constant RA arc until the fields
move by an hour angle equal to DECam's angular diameter, $\Theta_{\rm
  DECam}$, which means switching to a new set of fields after
approximately 9 minutes\footnote{the approximate time that it takes
  for $\Theta_{\rm DECam} \approx 2.2$ deg to transit the local
  meridian}. Thus, the exposure time and number of approximately
constant RA fields are related, as well as the exposure times plus
overheads with the total number of fields and epochs, by the following
approximate equations:
\begin{align}
  N_{\rm fields} &=  \frac{T_{\rm night}}{N_{\rm epochs} ~ (T_{\rm exp} + T_{\rm overhead})} \label{eq:Nfields}\\
  N_{\rm DEC}  &= \frac{\Theta_{\rm DECam}}{ 15 ~ {\rm deg / hr} ~ (T_{\rm exp} + T_{\rm overhead})}  \\
  N_{\rm RA} &=  \frac{N_{\rm fields}}{N_{\rm DEC}} 
\end{align}
where $N_{\rm fields}$ is the total number of fields to be visited,
$T_{\rm night}$ is the duration of the night, $N_{\rm epochs}$ is the
number of epochs per field, $T_{\rm exp}$ is the exposure time,
$T_{\rm overhead}$ is the overhead time (maximum between slew and
readout time), $N_{\rm RA}$ is the number of fields in the RA
direction in our grid of fields to observe and $N_{\rm DEC}$ is the
number of fields in the DEC direction in our grid of fields to
observe. The previous equations are only approximate and for a more
realistic selection of the optimal combination of fields to observe we
simulated the exact evolution of airmass and extinction with time for
different field configurations. Also note that because slew times are
given by the largest RA or DEC difference in the Blanco equatorial
mount we try to slew the telescopes diagonally in a RA--DEC grid.

To estimate the declinations that would minimize the combined
extinction from the Milky Way and the atmosphere we used Milky Way
extinction maps and the relation between atmospheric extinction and
airmass. This requires having defined the exact number of fields to
observe, which depend on the choices of the number of epochs per field
per night, $N_{\rm epochs}$, and the exposure time, $T_{\rm exp}$ (see
Equation~\ref{eq:Nfields}), which are given by the solution of the
survey design optimization problem.

\subsection{Figures of merit and simulations} \label{sec:simulations}

In this analysis we use two figures of merit: 1) the number of SNe
that would be detected at least once within the initial optical peak
caused by the shock breaking into the envelope of the RSG star and 2)
the number of SNe that would be detected at least twice within the
first day in rest--frame time after shock emergence. These two
quantities are related, but their comparison will give a sense of the
challenges associated with detecting the SBO optical peaks. Note that
for a transient candidate to be considered stationary (as opposed to
moving, e.g. asteroids) and to be detected in our online pipeline we
require a new event to be detected at least twice in image differences
in the same position in the sky.

In order to simulate how many SNe would be detected with these
restrictions we use the previously described light curves computed by
\citet{2011ApJS..193...20T} and \citet{2010ApJ...725..904N} (see
Figures~\ref{fig:SBO_LCs} and \ref{fig:day1_LCs}), assuming a
$\Lambda$--CDM cosmology with $\Omega_\Lambda = 0.73$, $\Omega_{\rm M}
= 0.27$ and $H_{\rm 0} = 71$ km s$^{-1}$ Mpc$^{-1}$ and the DECam
total efficiency. For the core collapse SN rate we use the cosmic star
formation rate density (SFRD) of \citet{2014ARA&A..52..415M}, asuming
a conversion efficiency between stellar mass and SNe, $\eta$, of
0.0091 $M_{\odot}^{-1}$ as measured by
\citet{2015ApJ...813...93S}. Note that this conversion efficiency does
not consider a large fraction of core collapse SNe missed by optical
surveys \citep{2012ApJ...756..111M}, so it is a conservative value for
our purposes. We also tried the SFR of \citet{2011ApJ...738..154H}, a
parametrization of \citet{2006ApJ...651..142H}, which resulted in
approximately 20\% more events.  Within a sufficiently large time
interval $T$ we simulate a sampling function, e.g. 4, 5 or 6 epochs
per night during several consecutive nights, and simulate a large set
of explosions to estimate the likelihood of the event being detected
at least once during the initial peak of a few hours seen in
Figure~\ref{fig:SBO_LCs} and the likelihood of the event being
detected at least twice during the first rest--frame day after shock
emergence. We record the distribution of detection ages to estimate
more precisely the likelihood of detecting an event with a given age.

We weight these results taking into account the initial mass function
(IMF) and normalizing their sum to 0.524 in order to reproduce the
observed fraction of SNe II among core collapse SN explosions found by
\citet{2011MNRAS.412.1441L}. In particular, we use a model weight
proportional to a Salpeter--like IMF integrated in the vicinity of the
model $M_{\rm ZAMS}$:
\begin{align} \label{eq:weights}
  w(M) \propto \int_{a_M}^{b_M} m^{-2.3} dm,
\end{align}
where $w(M)$ is the model weight associated with a zero age main
sequence mass $M$ and where $a_M$ and $b_M$ define the integration
mass interval for a given model, chosen to be either at the low mass
limit of 8.5 $M_{\rm \odot}$, equidistant between the masses of the
available models, or at either 16.5 or 30 $M_\odot$ at the high mass
limit. The low mass limit follows the low mass constraint from
\cite{2009MNRAS.395.1409S} and the high mass limit uses the high mass
constraint from the same work, but also considers the possibility that
RSG stars are surrounded by circumstellar material (CSM) before
explosion \citep[see e.g.][]{2012MNRAS.419.2054W, 2015ApJ...804...28G,
  2015MNRAS.451.2212G, 2016ApJ...818....3K, 2016arXiv160103261T,
  2016ApJ...820...23G}, possibly becoming hidden from progenitor
searches and allowing their masses to extend up to 30 $M_\odot$, which
is consistent with the 95\% confidence interval of
\cite{2012MNRAS.419.2054W}. These Salpeter--like IMF distributions
will be simply referred to as M16.5 and M30 distributions for the
upper mass limits of 16.5 and 30 $M_\odot$, respectively. 


 Additionally, we have calculated model weights which would be
 representative of the estimated masses of observed RSG stars in the
 Milky Way \citep{2005ApJ...628..973L} and the Magellanic clouds
 \citep{2006ApJ...645.1102L}, assuming that these RSG stars are at the
 base of the giant branch and counting the number of stars closer to
 the mass of a given model to compute the weigths. The model
 properties as well as the integral intervals for Salpeter--like IMF
 distributions and associated weights for all SN II IMF distributions
 are shown in Table~2.

\begin{deluxetable*}{c c c c | c c | c c | c c} 
  \centering \tablecaption{Explosion model properties from
    \citet{2011ApJS..193...20T} and model weights for the M16.5 and
    M30 distributions described in Section~\ref{sec:simulations}, as
    well as those representative of RSG stars in the Milky Way
    (RSG$_{\rm MW}$) and Magellanic clouds (RSG$_{\rm MC}$). The zero
    age main sequence (ZAMS), pre--SN radius, metallicity and
    explosion energy are shown, as well as the integration intervals
    in $M_\odot$ and model weights from
    Equation~\ref{eq:weights}. Note that the model weights sum is
    0.524, the observed fraction of SNe II in \citet{2011MNRAS.412.1441L}.}  
  \tablehead{
    \colhead{$M_{\rm ZAMS}$} & \colhead{R$_{\rm preSN}$} & \colhead{Z}
    & \colhead{E$_{\rm exp}$} & \colhead{$a^{\rm M16.5}_{M}$- $b^{\rm
        M16.5}_{M}$} & \colhead{$w^{\rm M16.5}$} & \colhead{$a^{\rm
        M30}_{M}$- $b^{\rm M30}_{M}$} & \colhead{$w^{\rm M30}$} &
    \colhead{$w^{\rm RSG_{MW}}$} & \colhead{$w^{\rm RSG_{MC}}$} }
  \startdata 13 $M_\odot$ & 564 $R_\odot$ & 0.02 & 1 foe & 8.5--14.0 & 0.433 & 8.5--14.0 & 0.310 & 0.126 & 0 \\
  15 $M_\odot$ & 507 $R_\odot$ & 0.02 & 1 foe & 14.0--16.5 & 0.091 & 14.0--17.5 & 0.086 & 0.154 & 0.091 \\
  20 $M_\odot$ & 795 $R_\odot$ & 0.02 & 1 foe & & & 17.5--22.5 & 0.071 & 0.181 & 0.370 \\
  25 $M_\odot$ & 1200 $R_\odot$ & 0.02 & 3 foe & & & 22.5--30.0 & 0.057 & 0.063 & 0.063 \\
  \enddata
\end{deluxetable*}

Assuming that all SNe II follow one of these explosion models we
estimate the number of events per unit redshift per field multiplying
the cosmic star formation rate density, $SFR(z)$; the conversion efficiency
between stars and SNe, $\eta$; the total time of our simulation, $T$;
the comoving volume per unit redshift bin per unit solid angle,
$\dfrac{dV}{dz d\Omega}(z)$; the field--of--view of DECam, $\Delta
\Omega$; and the detection probabilities in the first day, $P_{M}^{\rm
  d}(z)$, or in the initial peak, $P_{M}^{\rm p}(z)$, obtained for one
explosion model of mass $M$:
\begin{align}
  \frac{dn_{M}^{\rm d|p}}{dz} = \eta ~SFR(z)~ T ~ \frac{dV}{dz d\Omega}(z) ~\Delta \Omega ~P_{M}^{\rm d|p}(z).
\end{align}
The detection probabilities are obtained simulating large sets of
explosions in redshift bins of width 0.01 observed with the survey
cadence and depth. We assume an exponential host galaxy extinction
distribution $A_{\rm V}$ with a characteristic scale $\lambda_{\rm V}
= 0.187$ and $R_{\rm V} = 4.5$, as done in
\citet{2015ApJ...813...93S}. We count the number of events satisfying
the detection conditions and multiply this number by the efficiency
with which we can visually recognize SBO light curves as described in
Section~\ref{sec:surveycadence}.

For our survey design simulations we assumed a median seeing at CTIO
of 0.75'' in $r$--band \citep{2009PASP..121..922E}, converted to full
width half maximum (FWHM) values for different frequencies using the
relations found in the DECam exposure time calculator. Initially, we
computed the survey depth at the zenith using the available DECam
exposure time calculator (ETC) and then subtracted the excess
extinction with respect to the zenith at a given airmass using the
relations from \citet{1983MNRAS.204..347S}. After our survey was
completed we found that this significantly underestimated the effect
of airmass, and modified the available ETCs to also take into account
the effect of airmass on the FWHM, the atmospheric extinction and the
sky emission. We also used a 0.63'' DECam instrumental
FWHM\footnote{http://www.noao.edu/meetings/decam2015/abstract/Walker-Alistair}
and used a realistic evolution of the sky brightness\footnote{We
  compute the Moon's phase using PyEphem and interpolate the
  recommended values found in the DECam ETC} to compute 50\%
completeness magnitudes ($m_{\rm 50}$).

The number of detected SNe per field up to a given redshift will then
be the SN II IMF--weighted sum of the number of events per unit
redshift per field for the different explosion models of mass $M$:
\begin{align}
  N^{\rm d|p}(z) = \sum_{M} w(M) \int_0^z \frac{dn_{M}^{\rm d|p}}{dz} ~dz,
\end{align}

Different combinations of number of fields and number of epochs per
night per field are associated with different exposure times, which in
turn are associated with different limiting--magnitudes and number of
expected detections per night. In order to choose the best compromise
between these quantities keeping a cadence between 1.3 and 2 hours we
computed the total number of SNe that would be detected during SBO or
that would be detected twice during first rest--frame day for
different numbers of fields assuming 4, 5 and 6 epochs per night per
field in different bands.  We will show the results of these
simulations using empirically derived limiting--magnitudes in
Section~\ref{sec:results}.

\begin{figure*}[ht!]
 \centering
\includegraphics[scale=0.9]{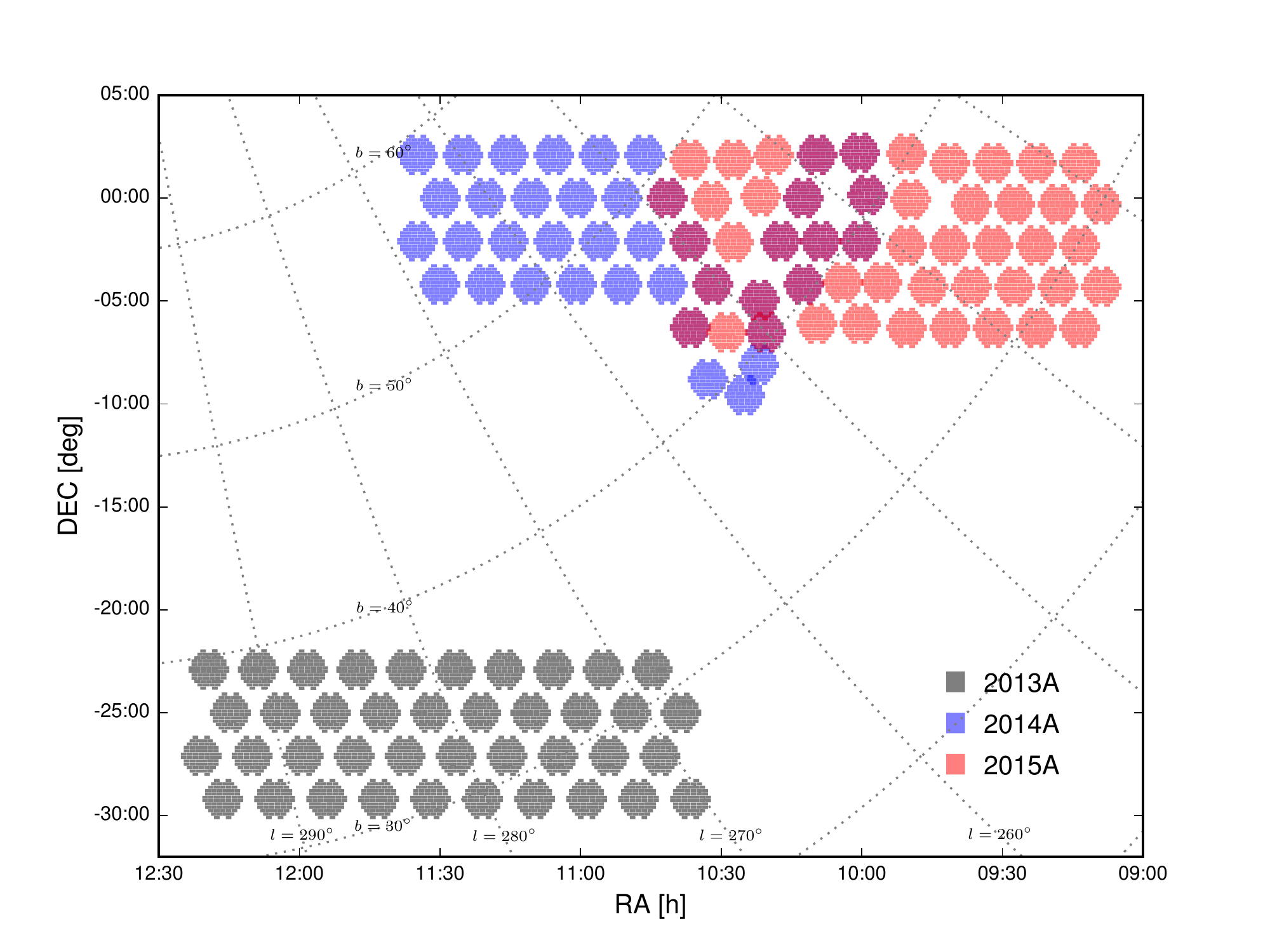} 
\caption{Spatial distribution of the fields observed in the 2013A,
  2014A and 2015A HiTS campaigns in equatorial (RA, DEC) and galactic
  ($b$, $l$) coordinates. See text for more details.}
\label{fig:fields}
\end{figure*}

For our 2013A pilot campaign we chose to observe in the $u$ band based
on the large temperatures during SBO predicted by our models. However,
in 2014A and 2015A we switched to the $g$ band after finding that this
band offered the best compromise between increased detector efficiency
and reduced SBO emission at larger wavelengths. Using these
simulations we chose 40 fields and 4 epochs per night per field for
both our pilot 2013A $u$ band campaign and real--time 2014A $g$ band
campaign. In 2015A we changed to 50 fields and 5 epochs per night per
field in order to explore faster cadences and to reduce the typical
limiting--magnitude of the survey, which facilitates follow up
observations with other telescopes. We will refer to the 2013, 2014
and 2015 HiTS survey campaigns as 13A, 14A and 15A, respectively. In
total, we were allocated four full nights in 13A, five full nights in
14A and six full nights in 15A, always near new Moon. In 15A we
obtained three additional half nights at later times for
multi--wavelength follow up. The final coordinates of the fields
observed in 13A, 14A and 15A are shown in Figure~\ref{fig:fields} and
their exact values are in the Appendix Section in Tables~A1, A2 and
A3. Note that the high cadence phase of observations were performed
near the new Moon: from Mar 12th to Mar 15th in 2013A (4 epochs per
field per night), from Mar 1st to Mar 5th in 2014A (4 epochs per field
per night) and from Feb 17th to Feb 22nd in 2015A (5 epochs per field
per night), which explains a variation in optimal central right
ascension by about two hours from 2013A to 2015A. Also note that in
order to minimize atmospheric extinction a declination close to the
latitude of CTIO ($-30^\circ$) is favoured, but in order to minimize
galactic extinction large galactic latitudes are favoured.

\section{DATA ANALYSIS} \label{sec:data}

The driving requirement for our data analysis strategy was the ability
to run in real--time and in a steady state, i.e. every exposure should
be fully processed in less than a given time interval, which should be
comparable to the exposure time plus overheads and which should be
much shorter than the typical cadence of the survey. This would allow
us to modify our observational strategy with DECam and to trigger
rapid follow--up observations with other telescopes \emph{during the
  same night of discovery} and obtain crucial progenitor information
only present during the first hours and days post--explosion
\citep[see e.g.][]{2014Natur.509..471G}. 

Given the data ingestion and computation rates imposed by the
real--time requirement, our data analysis pipeline had to be hosted in
a distributed memory type cluster with sufficiently fast connectivity
with the observatory, which is ideally offered by computational
facilities located in Chile today. Then, to have full control over the
pipeline's optimization we chose to write our own image subtraction
pipeline focusing on speed, trying to minimize input/output (I/O)
operations using the cluster node's shared memory for temporary data
storage, and avoiding redundant steps as much as possible.

\subsection{Pipeline outline} \label{sec:pipesteps}

The data reduction pipeline consists of a series of sequential steps
triggered in a distributed memory type cluster inmediately after every
observation, satisfying the following restrictions: 1)
non--parallelizable steps cannot take longer than the exposure time
plus overheads, e.g. data transfer, and 2) parallelizable steps cannot
take longer than the exposure time plus overheads times the ratio
between the total number of available cores and the product of the
number of parallel tasks and the number of cores per task.

The general structure of the pipeline is the following. Data is first
transferred to the distributed memory cluster to undergo an initial
pre--processing phase with data stored in the different node's shared
memory, including mosaic expansion, calibrations, cosmic ray removal
and indexing of the processed data. Then, the image subtraction
pipeline is triggered, again using the different node's shared memory,
and including registration, convolution, difference, photometry,
candidate classification and indexing of the processed data. Finally,
web visualization tools are updated with new candidates for human
inspection.

The pipeline was written mostly in
Python\footnote{http://www.python.org}, using Sun Grid Engine (in 14A)
and SLURM\footnote{http://slurm.schedmd.com} (in 15A) for job
distribution within the National Laboratory for High Performance
Computing (NLHPC)\footnote{http://www.nlhpc.cl} supercomputer.  We use
bash\footnote{http://www.gnu.org/software/bash/} for many I/O
operations, C for mosaic expansion (step \ref{step:expansion} below),
external software for pre--processing, cosmic ray removal and
catalogue generation (steps \ref{step:preprocessing},
\ref{step:cosmicrayremoval} and \ref{step:cataloguegeneration} below),
and Fortran 90\footnote{http://www.fortran90.org} subroutines
parallelized via OpenMP\footnote{http://openmp.org} and integrated
with Python using F2PY\footnote{http://www.f2py.com} \citep{P01} for
projection, convolution, difference object detection and photometry
(steps \ref{step:projection}, \ref{step:convolution} and
\ref{step:photometry} below). For the registration, convolution and
difference object detection we use a similar approach to existing
methods \citep[e.g.][]{1999ascl.soft09003A, 2015ascl.soft04004B},
although with important differences in the representation of the
convolution kernel and with our own Fortran routines for most of the
computationally demanding steps. The following sequential steps are
performed continuously during the night as data is acquired in the
telescope:

\begin{enumerate}

\item {\bf Data transfer}: \label{step:transfer} A bash script is left
  running on the telescope to upload raw DECam images from CTIO to the
  NLHPC (in 14A) or from La Serena to the NLHPC (in 15A) as soon as
  they arrive to the telescope's observer or La Serena preprocessing
  computer, respectively. The script looks for files which do not
  change in size after a lag of a few seconds and then transfers them
  to the NLHPC using \verb+rsync+\footnote{http://rsync.samba.org} in
  less than 10 seconds.

\item {\bf Image expansion}: \label{step:expansion} The standard
  pre--processing DECam pipeline expands mosaic data into their
  individual CCDs serially, which can add a significant lag to our
  real--time pipeline. Thus, raw DECam mosaic files were expanded into
  their individual CCD images using a parallel custom made program,
  \verb+pimcopy+, written in \verb+C+ and based on the CFITSIO library
  \citep{1999ASPC..172..487P}. This allows us to perform the image
  expansion in less than three seconds for one mosaic image, about 60
  times faster than a serial expansion.

\item {\bf Pre--processing}: \label{step:preprocessing} With the
  expanded files we run a pre--processing pipeline. In the 14A
  campaign we used a custom pre--processing pipeline which subtracted
  bias frames from the raw images, divided the resulting image by a
  flat field frame and computed inverse variance maps based on the
  Gaussian and Poisson noise components of the reduced images. The
  bias and flat field frames were obtained for every CCD combining
  several bias and flat field images observed during calibration time
  via median filtering and normalization. We also computed bad pixel
  masks for every CCD applying thresholds over the individual flat
  field images. However, we did not correct for CCD cross--talk
  effects which we found later to be important because of the
  appearance of \emph{ghost} images of bright stars. Therefore, in 15A
  we used a modified version of the DECam community pipeline (DCP) for
  the pre--processing stage, including electronic bias calibration,
  crosstalk correction, saturation masking, bad pixel masking and
  interpolation, bias calibration, linearity correction, flat field
  gain calibration, fringe pattern subtraction, bleed trail and edge
  bleed masking and interpolation.

\item {\bf Cosmic ray removal}: \label{step:cosmicrayremoval} We
  remove cosmic rays from the pre--reduced images using a modified
  version of the public code CRBLASTER \citep{2010PASP..122.1236M}
  using four cores per CCD via
  MPI\footnote{http://www.mpi-forum.org/docs/mpi-3.0/mpi30-report.pdf}. This
  code removes cosmic rays using Laplacian filter information as in
  \citet{2001PASP..113.1420V}. It takes about 20 seconds per CCD to
  remove cosmic rays, using 4 cores per CCD, or 248 cores in total.

\item {\bf Catalogue generation and reference
  image}: \label{step:cataloguegeneration} With the cosmic ray removed
  DECam images and previously obtained inverse variance maps we use
  SExtractor \citep{1996A&AS..117..393B} to derive source catalogues
  of the images in pixel coordinates. Given that we do not have
  previous reference images of the same area of the sky and that we
  are dealing with variability of hours, we use the first image of the
  sequence that has a relatively low airmass and good atmospheric
  conditions as our reference image for the relative and absolute
  astrometric solutions, to define a projection grid, for image
  difference and for the relative and absolute magnitude calibrations
  of the image differences. 

\item {\bf Astrometric solution}: \label{step:astrometricsolution}
  With the previously generated catalogues we match the brightest
  sources between the reference image catalogue and the new image
  catalogues in pixel coordinates. Given that we chose the same
  pointing between different epochs and that the 4~m Blanco telescope
  has an equatorial mount, there is only a small offset with almost no
  rotation between epochs. Then we solve for a quadratic
  transformation between the pixel coordinates of both images using
  least square minimization, which was found to give consistent
  results to using the astrometric solution to apply the projection
  and convolution. To obtain the astrometric solution of the reference
  image, where all the projections are made, we solve for the
  astrometric solution using the positions of known stars from the
  USNO catalogue \citep{2003AJ....125..984M} around our
  observations\footnote{In the latest version we fit the tangent point
    in the plane of the sky (CRVALs) and CCD (CRPIXs) as well as the
    scale and rotation matrix (CDs), using the CCD specific
    non--linear PV terms of DECam under the TPV WCS representation}.

\item {\bf Image projection}: \label{step:projection} Based on the
  previously derived transformations between the reference and new
  image in pixel coordinates we project the new image into the pixel
  grid of the reference image using a Lanzcos windowed sinc kernel
  with a kernel size of 2. This kernel has been found to be a good
  compromise in terms of reduction of aliasing, sharpness and minimal
  ringing \citep{TG90}. This takes about 10 seconds using four cores
  per CCD.

\item {\bf Kernel estimation}: \label{step:kernelestimation} In order
  to match the image point spread functions (PSFs) and scale of two
  images we use pixel based kernels with a non--uniform pixel size
  distribution and an approximately circular shape. This is a
  different kernel model than that used in
  \citet{2015ascl.soft04004B}, which uses an analytic basis to
  describe the kernel accross the CCD. Our model is pixel based
  \citep[c.f.][]{2016arXiv160801733M}, but with radially growing pixel
  sizes to reduce the number of free parameters per kernel from 625 to
  81, which has a regularization effect over the values of the
  outermost pixels. An example of the previously determined
  convolution kernels for a given CCD is shown in
  Figure~\ref{fig:kernel}. DECam is composed of 62 CCDs, each one of
  4k x 2k pixels. We divide every CCD in 18 (6 x 3) different regions
  with independently computed kernels. The size of the independent
  kernel regions was chosen by training kernels with only one bright
  star and then measuring the typical image subtraction residual
  change with distance for other test stars, such that the residual
  change at the kernel separation was comparable to the typical
  scatter of image subtraction residuals among different relatively
  bright stars. The kernels are derived via least square minimization
  of multiple pairs of stars taken from the same region of the CCD,
  selected for being isolated, bright and within the linear response
  regime of the CCD. The kernels are defined so that the smaller full
  width half maximum (FWHM) image is convolved into the larger FWHM
  image, in order to avoid effective deconvolutions.

\begin{figure}[!ht]
\centering
\includegraphics[scale=0.32]{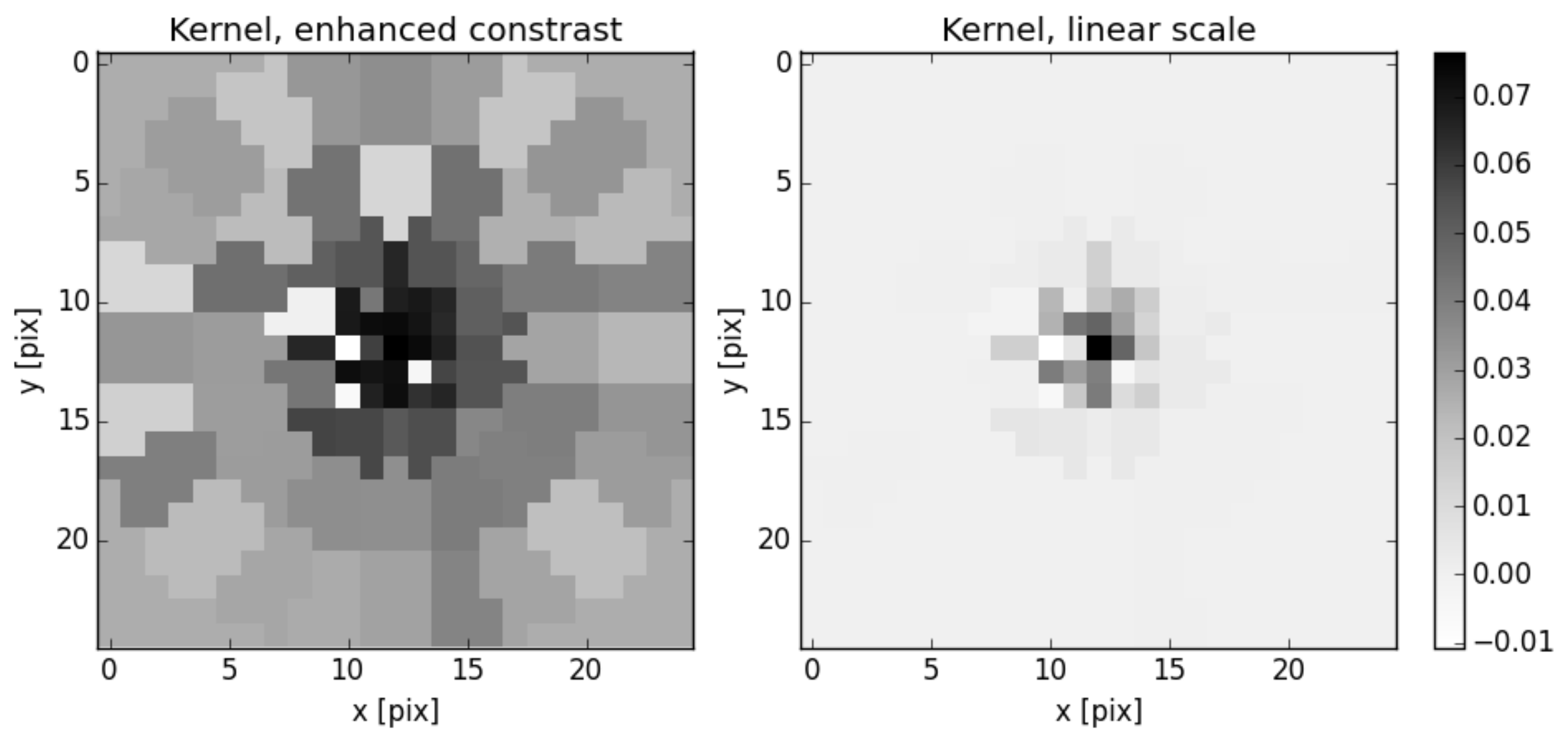} 
\caption{Example kernel used for the PSF matching convolution in a
  section of a pair of images, in non--linear scale to enhance the
  contrast at low pixel values (left) and in linear scale
  (right). }
\label{fig:kernel}
\end{figure}

\item {\bf Image convolution and difference}: \label{step:convolution}
  We use the previously derived kernels to convolve the corresponding
  image and region of the CCD where it was defined. This changes both
  the shape and scale of the PSF of the convolved image. We then
  subtract the science and reference images, where one of them would
  have been convolved, to obtain difference and difference variance
  images in the units of the image which has not been convolved. This
  takes less than 30 seconds using four cores per CCD.

\item {\bf Image difference photometry}: \label{step:photometry} Using
  the difference and difference variance images we then perform an
  optimal photometry calculation \citep{1998MNRAS.296..339N} over the
  entire image assuming that the PSF is centered in every pixel. This
  is achieved in less than 10 seconds using four cores per CCD,
  producing a PSF integrated photometry image centered in every pixel
  and an associated PSF integrated photometry variance image, which
  are scaled up or down into the units of the reference image. We then
  select those pixels with signal to noise ratio (SNR) greater than
  five to obtain candidate image stamps, using the optimal photometry
  flux and variance, which are fed into a machine learning
  classifier. Computing SNRs for measured fluxes centered in every
  pixel of the image allows us to perform normality tests over the SNR
  distribution for real--time quality control.

\item {\bf Candidate filtering}: \label{step:candidatefiltering} Every
  candidate is used to create candidate image stamps whose
  dimensionality is reduced with custom--designed features that are
  input for a random forest classifier \citep[e.g.][]{Breiman2001,
    2007ApJ...665.1246B, 2015AJ....150...82G} to be classified as
  either real or bogus. Because the training set was based on the 13A
  campaign (in the $u$ band) to be used in the 14A/15A campaigns (in
  the $g$ band), a key principle during the feature engineering
  process was to use as many dimensionless and scaleless quantities as
  possible.
  
  In order not to be dominated by unknown moving objects,
  e.g. asteroids, we consider that true non--moving transients are
  those that appear twice within a distance consistent with the
  astrometric errors and when at least one of the difference fluxes is
  positive with respect to the reference frame. A repetition of the
  candidate rules out moving transients, e.g. asteroids (to be
  presented in a separate publication), which are dominant among the
  candidates we detect, but only if it has at least one positive
  difference flux with respect to the reference frame. This is because
  a moving transient in the reference image should be present in all
  difference images with a negative difference flux with respect to
  the reference.

\item {\bf Light curve generation}: \label{step:lightcurvegeneration}
  Once a candidate has passed all the previous tests a light curve is
  generated. By default, the light curve contains only those
  difference fluxes of candidates that were classified as true
  transients by the random forest classifier. Optionally, the light
  curve can contain the difference fluxes from all the available
  epochs, even when the candidate was not selected by its SNR or if it
  was not selected by the machine learning classifier. This can be
  computed revisiting the difference images and doing a forced
  photometry (fixing the PSF shape and position) using the optimal
  photometry method, which may take significantly longer times than
  all the previous steps because of the many more I/O operations
  required. For this reason, we revisit all the epochs only for a few
  visually selected candidates during the night, and for all the
  candidates classified as real non--moving transients during the day.

\item{\bf Webpage visualization}: \label{step:webpagegeneration} As
  soon as a candidate is classified as a true candidate it will be
  linked to a website that displays in visual form the location of all
  the candidates within the CCD. Candidates that are repeated and that
  have at least one positive difference flux with respect to the
  reference frame are marked with different colors to aid the visual
  inspection (using a combination of PHP\footnote{http://www.php.net},
  JavaScript\footnote{http://www.javascript.com} and
  Highcharts\footnote{http://www.highcharts.com}). The positions of
  known stars, galaxies, moving objects and variable stars are also
  shown with different colors. We mark those candidates that are found
  to be consistent with having periodic light curves, using a
  Lomb-Scargle periodogram analysis \citep{1976Ap&SS..39..447L,
    1982ApJ...263..835S}, to separate fast periodic stars from fast
  transients. All fields for a given campaign are displayed in a
  single map, with about three new candidates appearing at the
  location of the last field visited after every observation, giving
  higher visibility to those candidates which have not been inspected
  in order to aid with the process. The visual inspection of all
  fields and CCDs can be done in about one hour by a single person,
  but it can be easily accomplished in real--time with these
  visualization tools.

\item {\bf Follow up}: If a candidate is designated as a true
  non--moving transient by the previous tests it will be followed--up
  only after going through the visual inspection test. We then decide
  whether to change DECam's observation plan, to obtain more
  observations in the same or other bands, or to trigger photometric
  or spectroscopic observations with other telescopes. Although we had
  the capability for triggering spectroscopic follow--up in a time
  comparable to about twice the cadence, no clear SBO candidates were
  found. Instead, during the last night of the 14A/15A high cadence
  run we changed our observation to include multiple wavelengths
  and/or triggered spectroscopic observations towards some of the
  rising SNe.

\end{enumerate}

\begin{figure*}[!ht]
 \centering
\includegraphics[width=0.6\textwidth]{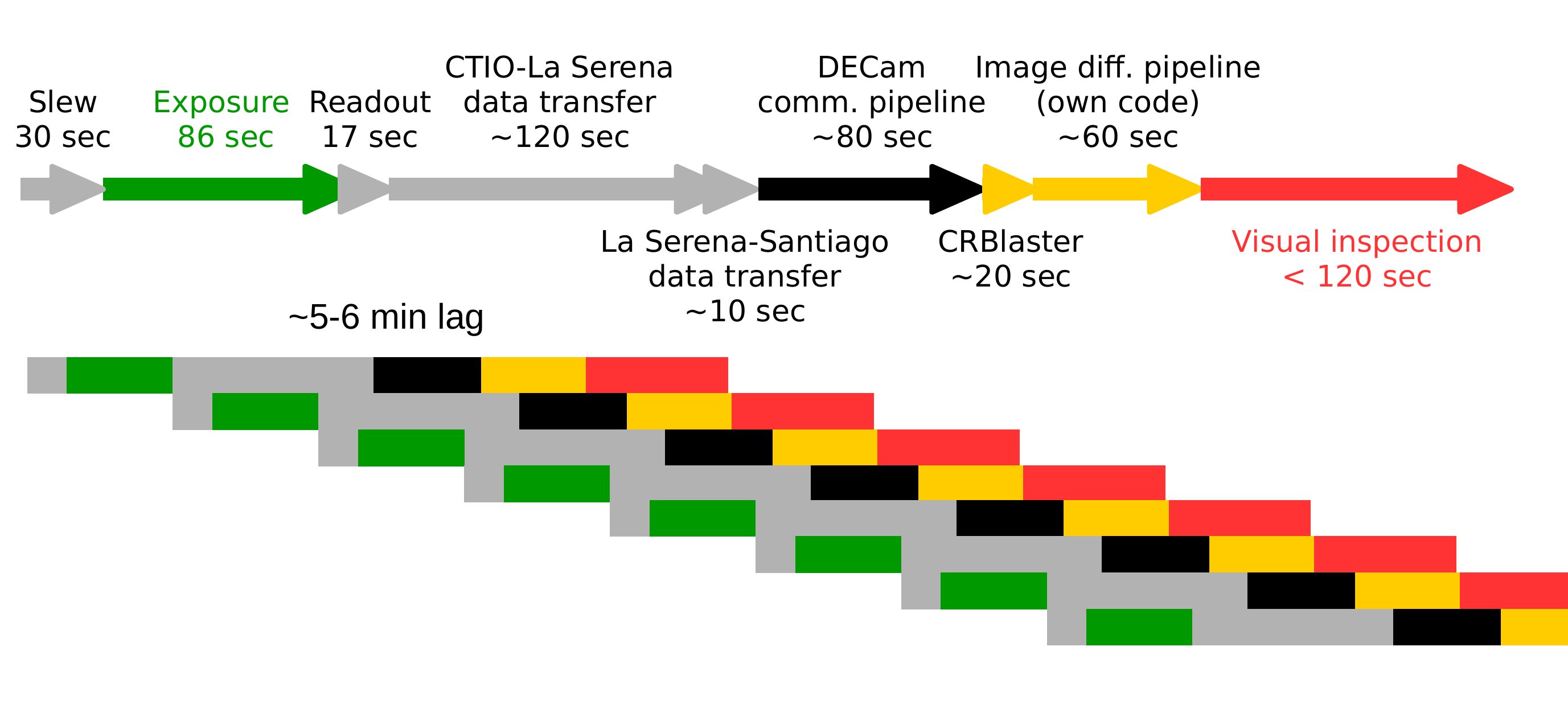} 
\caption{Schematic representation of the data analysis pipeline from
  telescope slew to candidate visualization for the 2015 campaign. The
  top arrows indicate the pipeline steps described in the text and
  their approximate execution times. The bottom blocks represent the
  previous steps when the pipeline is executed in real--time during
  the fast cadence phase of the survey, resulting in an online
  processing with a lag of less than 6 minutes between shutter close
  and end of candidate visualization for every observation. }
\label{fig:pipeline}
\end{figure*}

In total, we processed more than $10^{12}$ pixels at a maximum rate of
4.5 Mpix/sec, or 40 Mbps, generating about $10^8$ significant image
subtraction candidates (with a SNR greater than 5), of which about
$10^6$ were classified as not bogus. Less than $10^4$ of these
candidates were selected by the classifier twice in the same position
in the sky, which were then visually inspected in a stream of about 3
new candidates per minute. The total lag time between shutter closing
and end of candidate visualization for a given field was an
approximately constant time of about 5 to 6 minutes taking into
account all the previous steps. A visualization of the real--time
pipeline steps is shown in Figure~\ref{fig:pipeline} for the 15A
campaign, which had the shortest expsure times of the three available
campaigns. In this figure the yellow arrow labeled \emph{Image
  diff. pipeline} contains steps~\ref{step:cataloguegeneration} to
\ref{step:lightcurvegeneration}. This figure shows why the visual
inspection should be done as fast as the telescope slew plus exposure
and that the DECam community pipeline is normally run in parallel with
the image difference pipeline from a previous exposure.

\section{RESULTS} \label{sec:results}

\subsection{Survey depth} \label{sec:depth}

In this section we discuss the empirically determined depth of the
survey, its relation to airmass and other observational variables and
how it compares to our survey design assumptions.

\begin{figure}[!ht]
\centering
\vbox{
\includegraphics[scale=0.25]{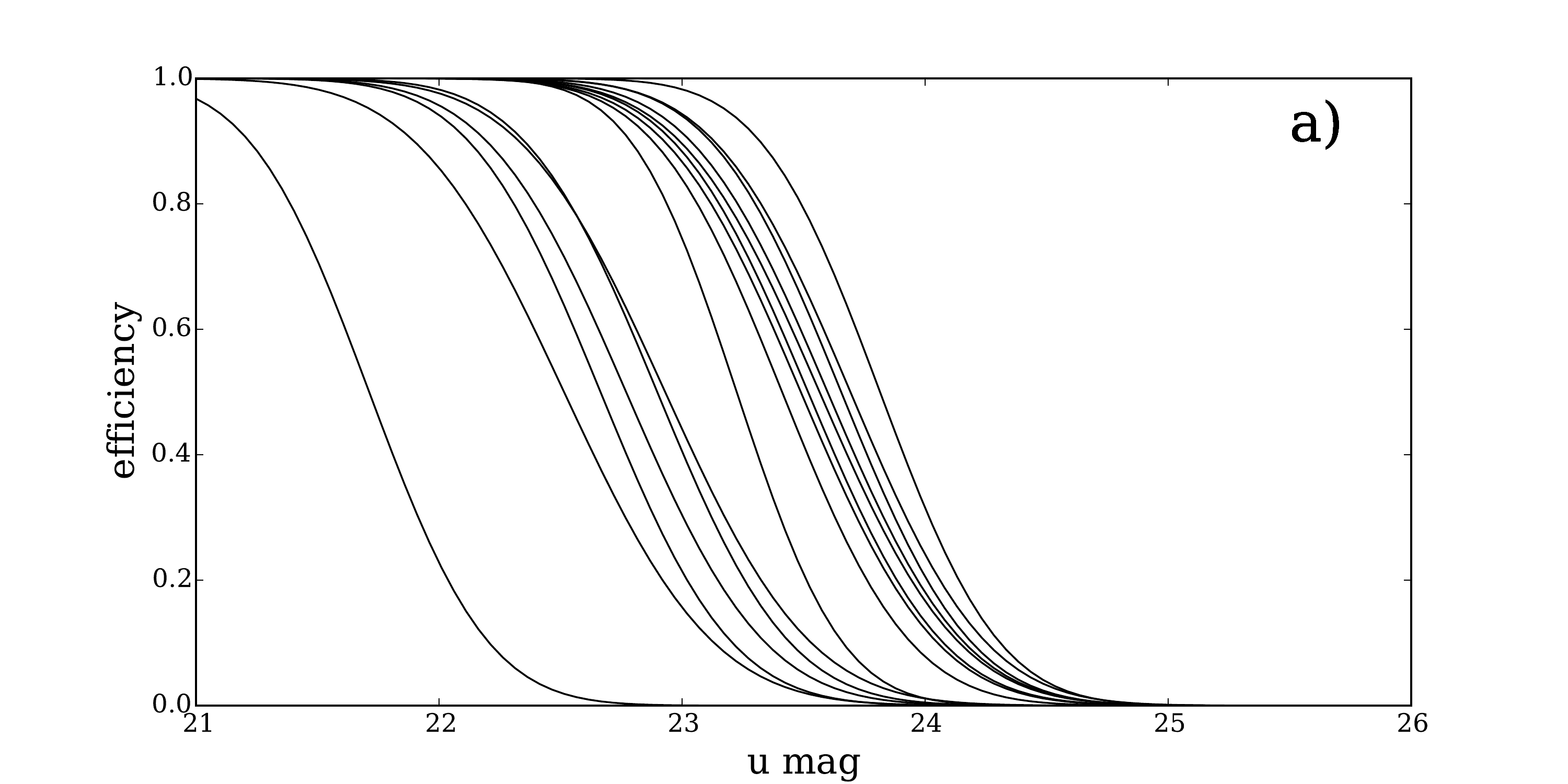} 
\includegraphics[scale=0.25]{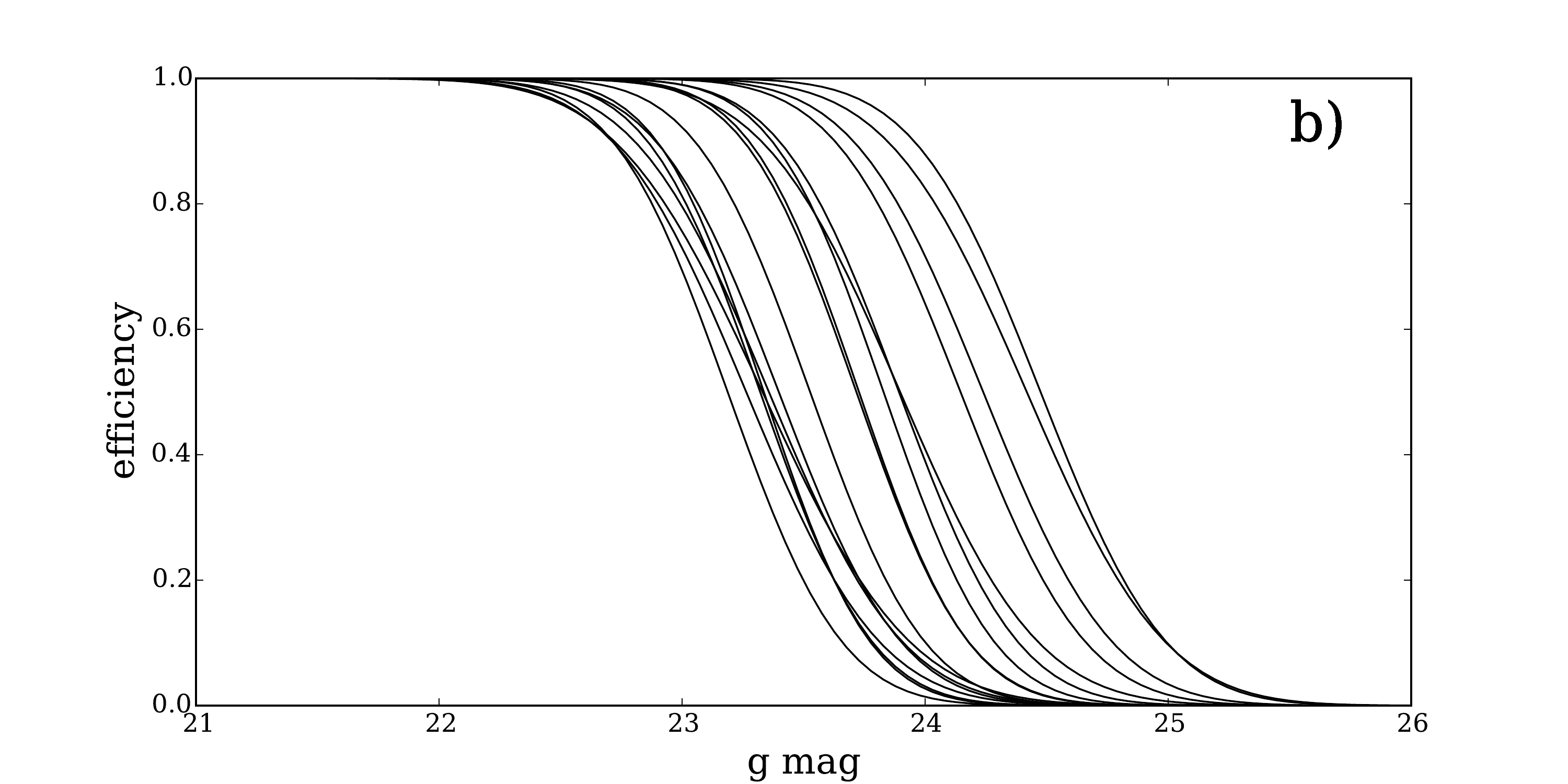} 
\includegraphics[scale=0.25]{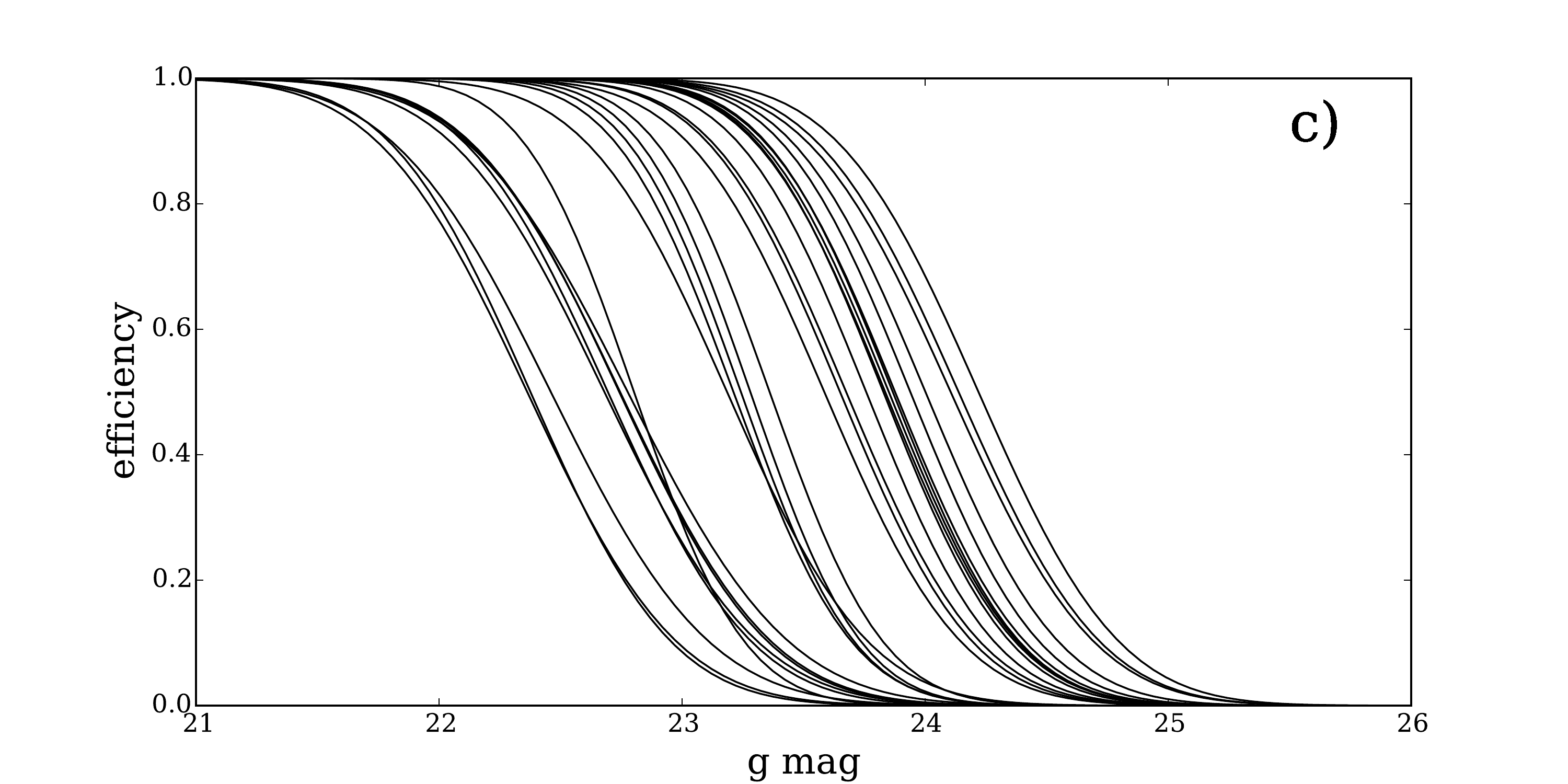} 
}
\caption{Best--fitting efficiency vs magnitude relations for a set of
  observations in the 13A {\bf (a)}, 14A {\bf (b)} and 15A {\bf (c)}
  campaigns. The differences between observed relations are due to
  different airmasses and changing environmental conditions at which
  the observations were performed.}
\label{fig:efficiencies}
\end{figure}

\begin{figure}[!ht]
\centering
\includegraphics[scale=0.42]{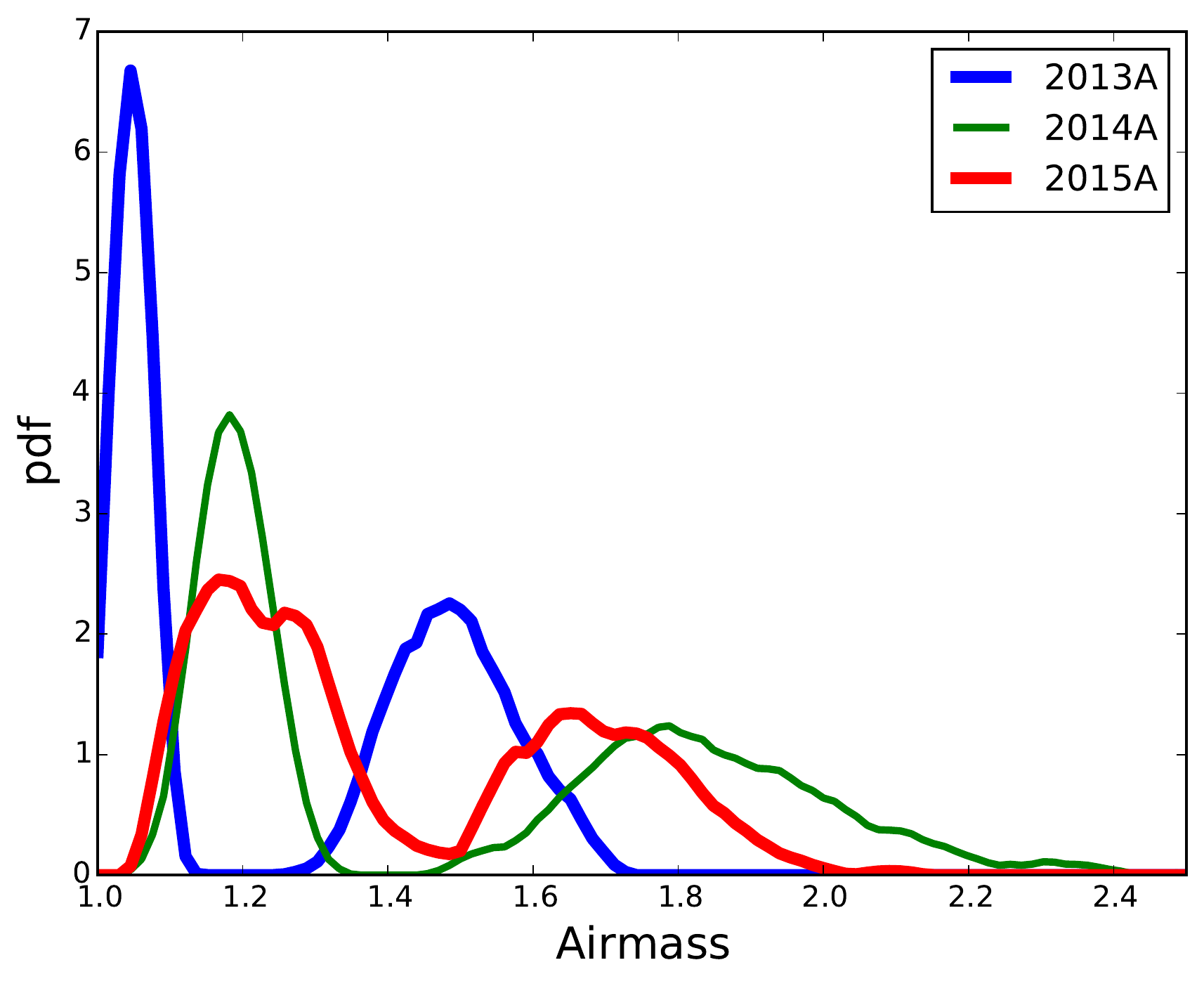} 
\caption{Airmass distribution of the 13A, 14A and 15A HiTS
  campaigns. In both the 13A and 14A campaigns we observed 4
  epochs per night per field, two at lower airmasses and two at larger
  airmasses, while in the 15A campaign we observed 5 epochs per
  night per field (using shorter exposure times), three at lower
  airmasses and two at larger airmasses.}
\label{fig:airmasses}
\end{figure}

The depth os the survey in each of these campaigns were measures as an
efficiency (probability of detection) vs magnitude for every observed
epoch separately. This can be used to better estimate the expected
number of events that should have been detected in the entire series
of observations of a given field. The probability of detection at
different magnitudes for a given set of representative observations is
computed by building deep image stacks, co--adding selected epochs of
a series of observations of the same field, and then measuring the
fraction of real objects that were detected in single epoch
observations relative to those in the deep stacks for a given
magnitude range. Assuming that in the deep stacks all the true sources
near the limiting--magnitude of the individual images are detected, we
measure the efficiency of detections in single epoch observations as a
function of magnitude taking into account all available CCDs
simultaneously. We then fit the observed fraction with the following
function:
\begin{align} \label{eq:eff}
P(m) = \frac{1}{2} \biggr[1 + {\rm erf} \biggl( - \dfrac{m -
    m_{50}}{\Delta m_{50}} \biggr) \biggr],
\end{align}
where $P(m)$ is the probability of detection of a stellar--like source
at a given magnitude, $m$; ${\rm erf}$ is the error
function\footnote{${\rm erf(x)} = \frac{2}{\pi} \int_0^x e^{-t^2}
  dt$}; $m_{50}$ is the best--fitting 50\% completeness magnitude; and
$\Delta m_{50}$ is a scale parameter approximately equal to half the
width of the transition region to low efficiencies. We tried several
analytic expressions, but we found Equation~\ref{eq:eff}, using the
error function, to better resemble the observed transition in
efficiency. The best--fitting analytic approximation of the
efficiencies for all the CCDs of one of the fields observed in 13A,
14A and 15A is shown in Figure~\ref{fig:efficiencies}. The large
spread in 50\% completeness magnitudes is due to the airmass
variations required in our observing strategy, shown in
Figure~\ref{fig:airmasses}, but also due to bad observing conditions
in some epochs. $\Delta m_{50}$ best-fitting values were typically
$0.6 \pm 0.1$ mag.

\begin{figure*}[!ht] 
  \centering
    \includegraphics[scale = 0.4]{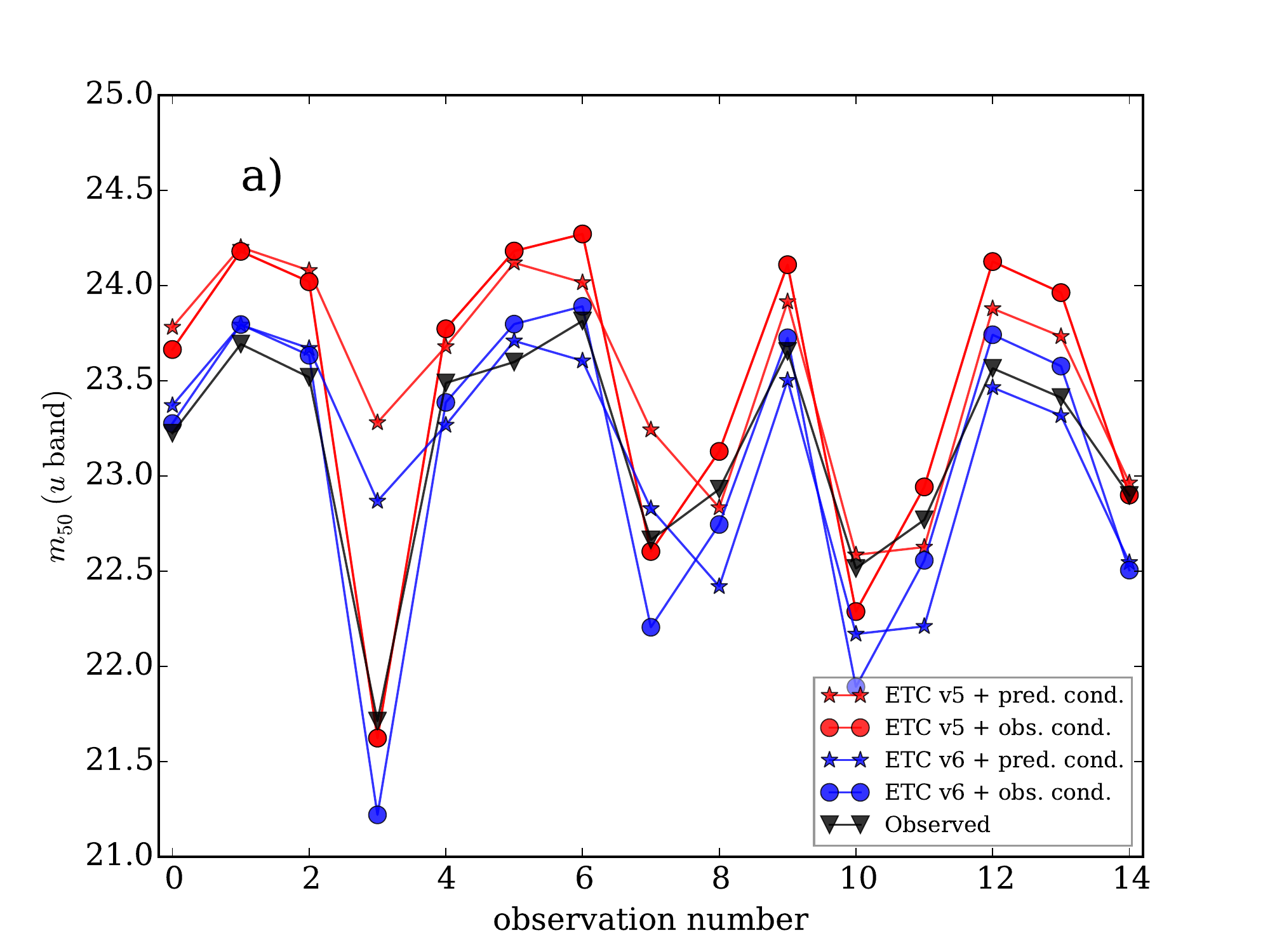} 
  \hbox{
    \includegraphics[scale = 0.4]{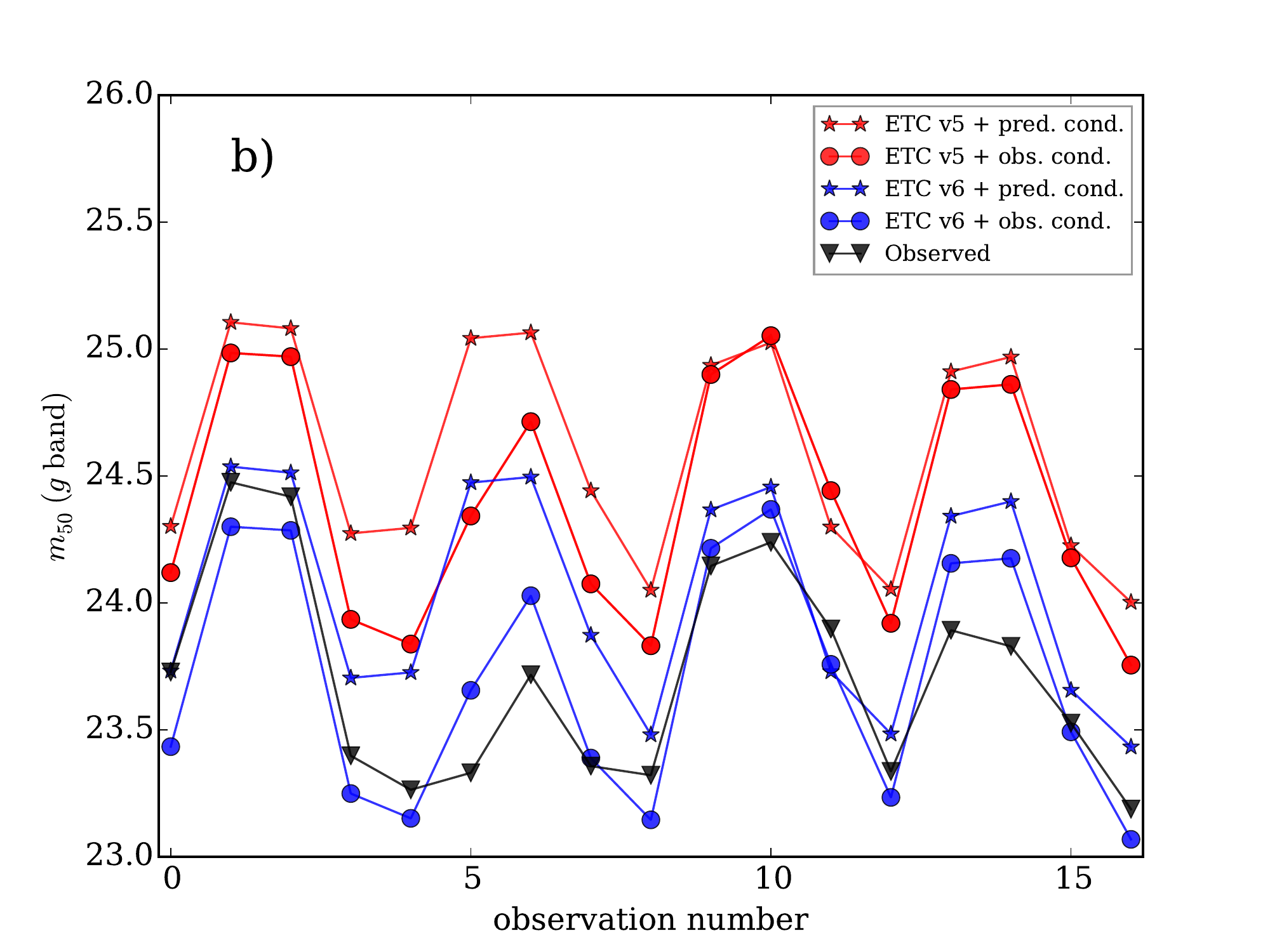} 
    \includegraphics[scale = 0.4]{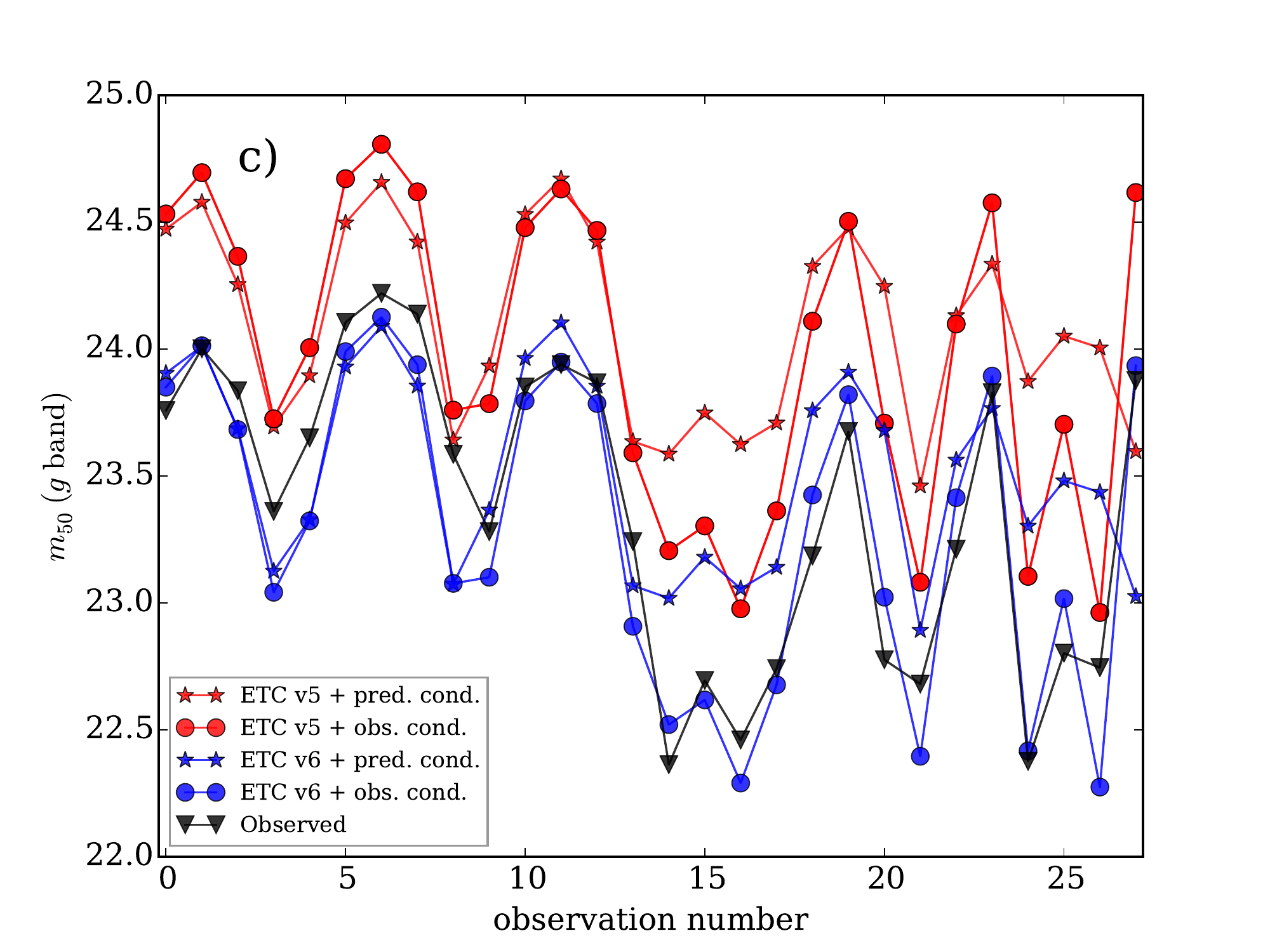} 
  }
  \caption{Best--fitting 50\% completeness magnitudes ($m_{\rm 50}$)
    for the 13A {\bf (a)}, 14A {\bf (b)} and 15A {\bf (c)} survey
    campaigns and the 5--$\sigma$ limiting--magnitudes reported by the
    ETCs v5 and v6 modified to include the effect of airmass as a
    function of observation number. We use observation number instead
    of time to avoid having data points too clustered to accommodate
    the daily gaps. See text for more details.}
\label{fig:m50vsETC}
\end{figure*}

In Figure~\ref{fig:m50vsETC} we compare the best--fitting 50\%
completeness magnitudes ($m_{\rm 50}$) and the 5--$\sigma$
limiting--magnitudes reported by the DECam exposure time calculators
versions 5 and 6 (ETC v5 and v6, respectively) modified to include the
effect of airmass. Note that the ETC v5 is included because it was the
latest available ETC at the time when the 15A semester observations
were performed. We show the 5--$\sigma$ limiting--magnitudes using the
modified ETC v5 with the predicted FWHM and sky brightness (\emph{ETC
  v5 + pred. cond.}), the modified ETC v5 with the observed FWHM and
sky brightness (\emph{ETC v5 + obs. cond.}), the modified ETC v6 with
the predicted FWHM and sky brightness (\emph{ETC v6 + pred. cond.}),
the modified ETC v6 with the observed FWHM and sky brightness
(\emph{ETC v6 + obs. cond.}) as well as the empirical 50\%
completeness magnitude (\emph{Observed}). Note that in the 15A
campaign the 4$^{th}$ and 5$^{th}$ nights were strongly affected by
cloud cover and resulted in only three epochs in total (observation
numbers 14 to 16) and that the 15A campaign contained follow up
observations starting at observation number 22. In all plots the
measured 50\% completeness magnitude matches better the 5--$\sigma$
limiting--magnitude produced with the modified ETC v6 using the
empirical FWHM.  We found that there is a difference of approximately
0.6 magnitudes between the ETC v5 and v6, but also that there were
FWHM variations beyond what is expected from airmass variations, which
are also required to explain some of the $m_{\rm 50}$ variations.

As mentioned before, initially we used the relation between airmass
and extinction from \citet{1983MNRAS.204..347S} to calculate an excess
extinction with respect to the limiting magnitude at the zenith, which
is given by the original ETC, and use this information to derive the
limiting magnitude at a given airmass. However, this led to a
significant overerestimation of the limiting--magnitudes. Instead, in
the modified version of the ETCs we assume that the FWHM scales as
$\cos(z)^{-3/5}$, where $z$ is the zenithal angle, that the sky
emission scales linearly with airmass and that the atmospheric
extinction increases exponentially with airmass to derive the limiting
magnitude at a given airmass. A Python version of our modified ETC can
be found in \url{https://github.com/fforster/HiTS-public}.

\begin{figure*}[]
  \centering
  \includegraphics[scale = 0.45]{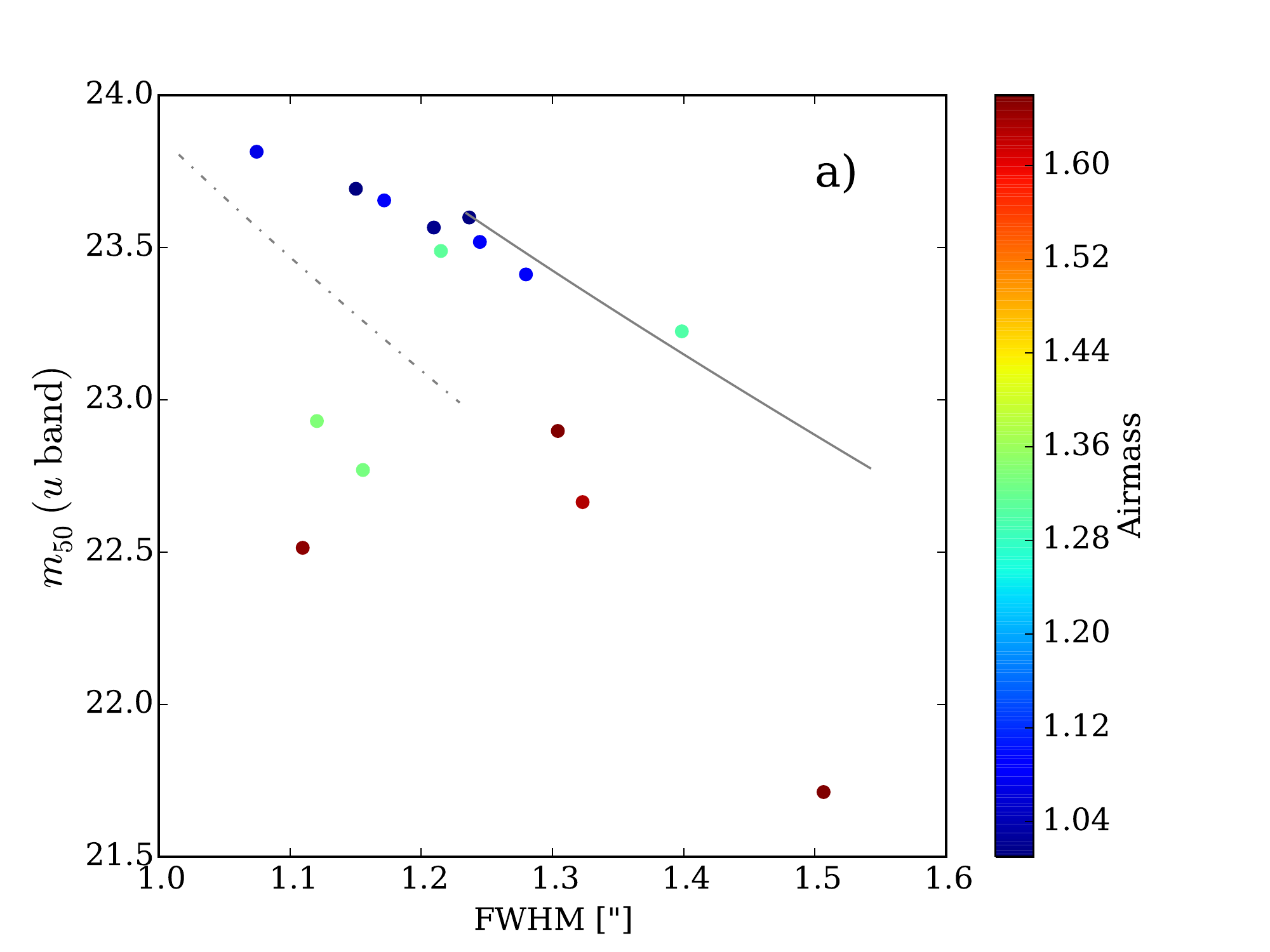} 
  \hbox{
    \includegraphics[scale = 0.45]{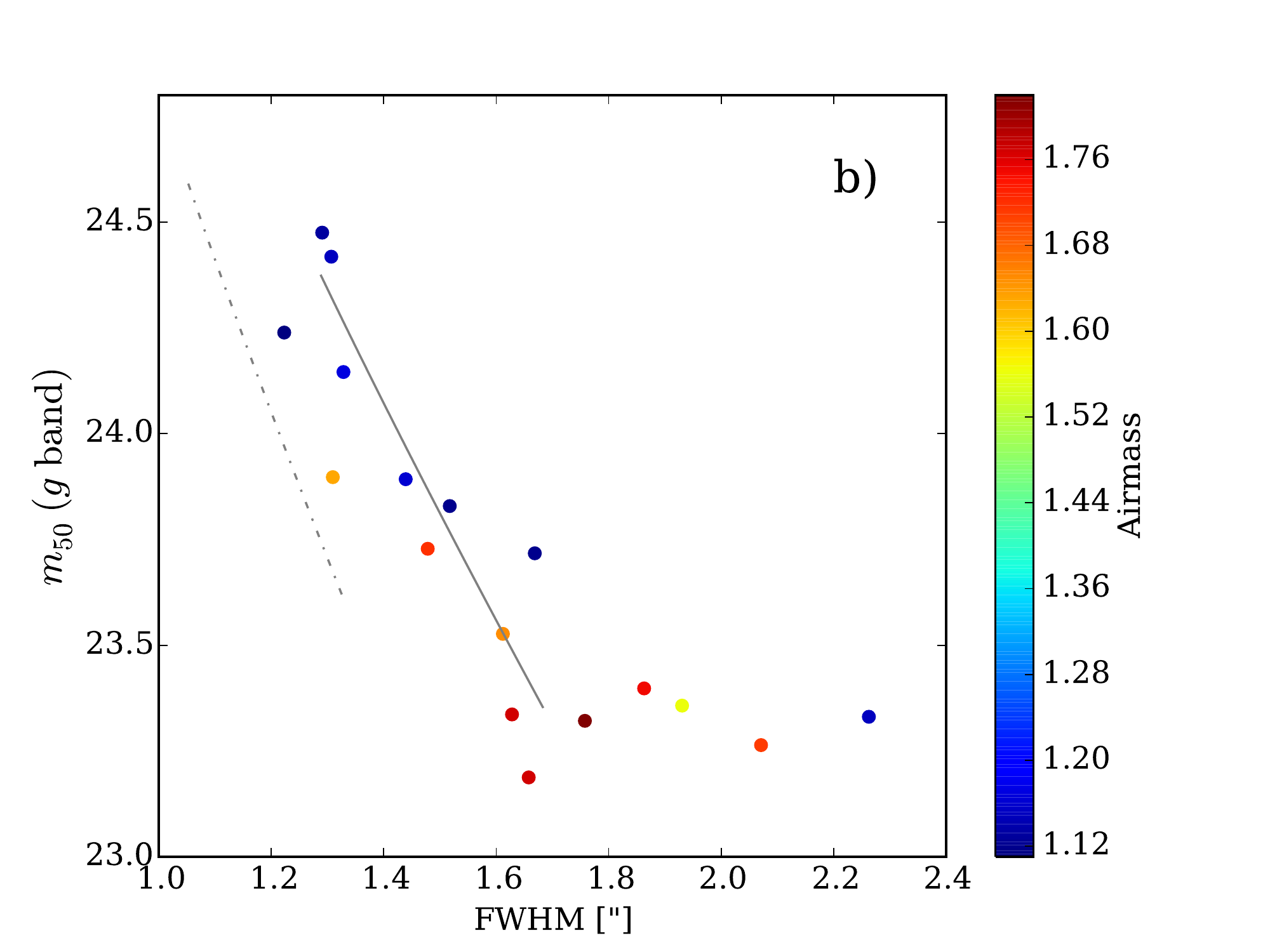} 
    \includegraphics[scale = 0.45]{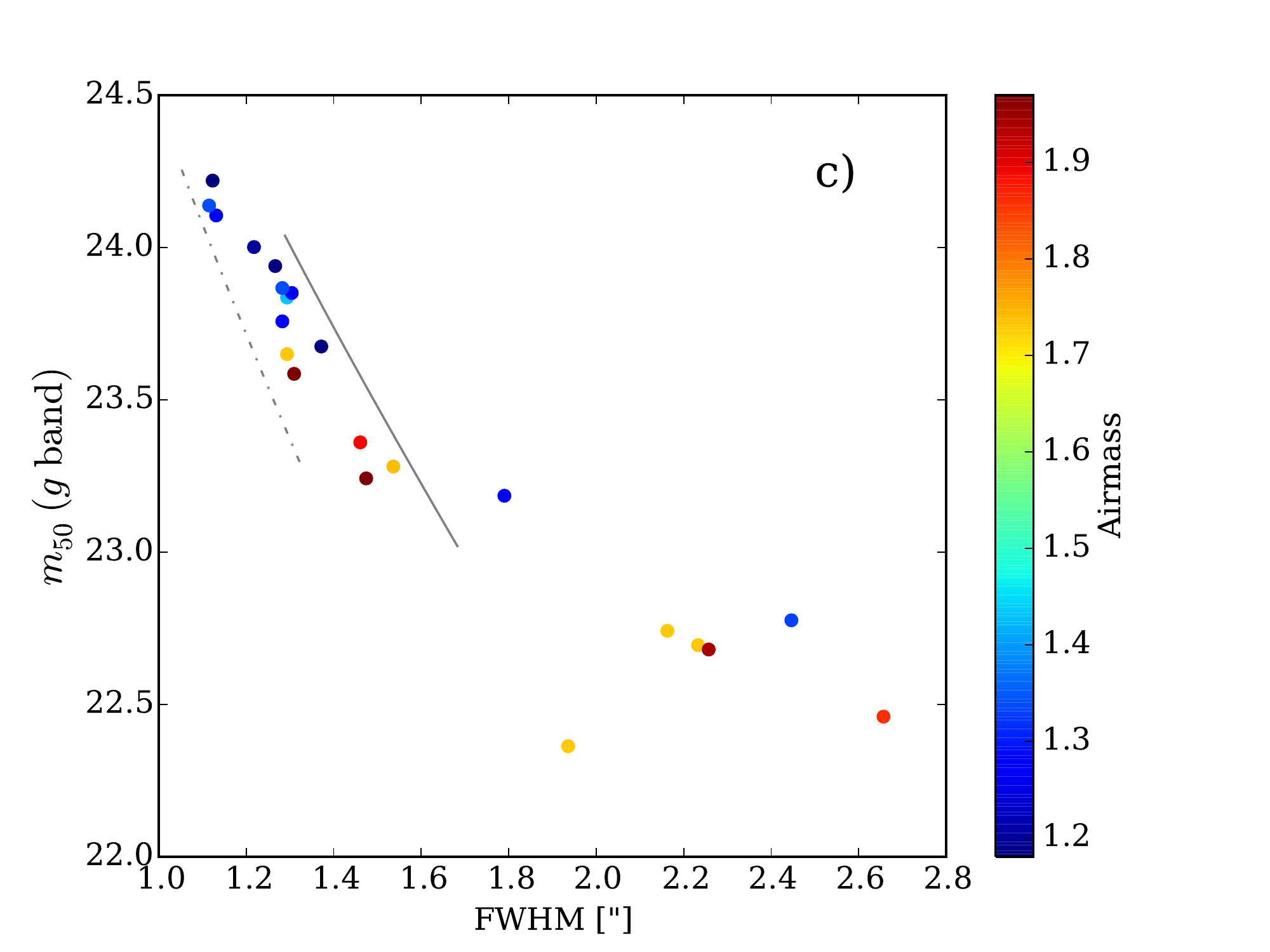} 
  }
  \caption{Relation between the measured 50\% completeness magnitude
    ($m_{\rm 50}$), the FWHM and the airmass for mosaic averaged
    pointings in typical fields of the 13A {\bf (a)}, 14A {\bf (b)}
    and 15A {\bf (c)} survey campaigns. Gray lines show the expected
    relation for different FWHM at zenith. See text for mode details.}
\label{fig:m50-FWHM-airmass}
\end{figure*}

The relation between $m_{\rm 50}$, FWHM and airmass is shown in
Figure~\ref{fig:m50-FWHM-airmass}.  The expected relation between
$m_{\rm 50}$ and FWHM for constant seeings of 0.75'' (dot--dashed gray
line) and 1.0'' (continuous gray line) in the $r$ band are shown,
scaled to the $u$ or $g$ bands and using a range of airmasses (between
1.0 and 1.6 for 2013A and between 1.1 and 1.9 for 14A and 15A) to
derive the FWHMs, where we also assume an additional 0.63'' DECam
instrumental seeing.  We use a fixed sky brightness of 22.6
mag/arcsec$^2$ in the $u$ band for 13A and 22 mag/arcsec$^2$ in the
$g$ band for 14A and 15A, scaled linearly with airmass (in physical
units) with the modified ETC v6. With the modified ETCs we expect that
for the typical range of airmasses the limiting--magnitudes can vary
as much as 1.1 mag.  Thus, the upper--left areas of
Figure~\ref{fig:m50-FWHM-airmass} can be explained by an airmass
effect, but the bottom--right areas are mostly due to changing
observing conditions, including the appearance of clouds in some of
the observing nights.

The previous efficiencies are based on single epoch measurements on
full DECam mosaics, but our detections are based on image
differences. To account for the additional loss of efficiency due to
image differencing, we injected artificial stars and studied their
signal to noise ratio (SNR) at injection and after recovery,
concluding that the effect was typically a decrease in the SNR of up
to $\sqrt{2}$, which would be expected when subtracting two images
with the same noise level, or a shift in the efficiency vs magnitude
relation of up to $2.5 ~ \log_{10} (\sqrt{2}) \approx 0.38$
magnitudes. This is because we did not use deep templates for image
subtraction, but used the first observation at a given band that
approached the lowest airmass values achievable (about 1.05 in 13A and
1.2 in 14A/15A, see Figure~\ref{fig:airmasses}) of the first night
which had photometric conditions (2$^{nd}$ observation in the main
band for the three campaigns), effectively amplifying the noise by up
to $\sqrt{2}$ since the images are dominated by sky shot
noise. Therefore, we shifted our efficiency--magnitude relations by
0.38 magnitudes to account for the effect of image differences, but
also used the original unshifted relation as best--case scenario in
what follows.

Overall, there are four relevant factors which made the survey depth
shallower than expected: 1) the exposure time calculator (ETC)
available when designing the survey (v5) overestimated the
limiting--magnitudes by approximately 0.6 magnitudes with respect to
the latest ETC version (v6), which matches well our empirical 50\%
completeness magnitudes; 2) the effect of airmass was larger than
expected, with a variation of up to 1.1 magnitudes in the 50\%
completeness magnitudes instead of less than 0.3 magnitudes using the
relations from \citet{1983MNRAS.204..347S}; 3) the observational
conditions were not always good, with nights that had significantly
larger than expected seeing values or even clouds; and 4) the
difference imaging could lead to a loss in depth of up to 0.38
magnitudes with respect to the individual epoch measured depths. The
combination of all the previous effects can be seen in the large
magnitude differences seen in Figure~\ref{fig:m50vsETC}.

A practical description of the main observational parameters for the
13A, 14A and 15A observational campaigns is shown in Table~3. Note
that for the 15A campaign we had three DECam half nights separated by
two, five and twenty days from the end of the high cadence phase,
where we also observed selected fields in the $r$ and $i$ bands.

\begin{deluxetable*}{c c c c c c c c c c}
  \centering \tablecaption{HiTS survey description summary}
  \tablehead{
    \colhead{Semester}
    & \colhead{Band} 
    & \colhead{Area [deg$^2$]}
    & \colhead{\# nights}
    & \colhead{\# fields}
    & \colhead{\# epochs/night}
    & \colhead{Exposure [s]}
    & \colhead{Cadence [hr]}
    & \colhead{Airmass}
    & \colhead{Typical $m_{\rm 50}$}}
  \startdata
  13A & $u$ & 120 & 4 & 40 & 4 & 173 & 2 & 1.0--1.6 & 23--24 \\
  14A & $g$ & 120 & 5 & 40 & 4 & 160 & 2 & 1.1--2.1 & 23.5--24.5 \\
  15A & $g$ & 150 & 6 & 50 & 5 & 87 & 1.6 & 1.1--1.9 & 23--24.5
  \enddata
\end{deluxetable*}

\subsection{Number of detections}

The lower than expected survey depths obtained in
Section~\ref{sec:depth} (at least 0.6 magnitudes of difference) should
have an important effect on the predicted number of shock breakout
(SBO) events, which means that our initial assumptions during survey
design had to be corrected. Thus, we have recomputed the expected
number of events for different observational strategies using
different assumptions about the survey depth. We use both the modified
ETCs v5 and v6 for different number of fields per night, assuming no
airmass evolution or a realistic airmass evolution given a typical
declination and 4 or 5 epochs per field per night for 13A and 14A or
15A, respectively. We also test for the effect of the image difference
process by assuming that it results in a loss of depth of 0.38
magnitudes or assuming that it results in no loss in depth.

\begin{figure*}[!ht] 
  \centering
  \includegraphics[scale = 0.55]{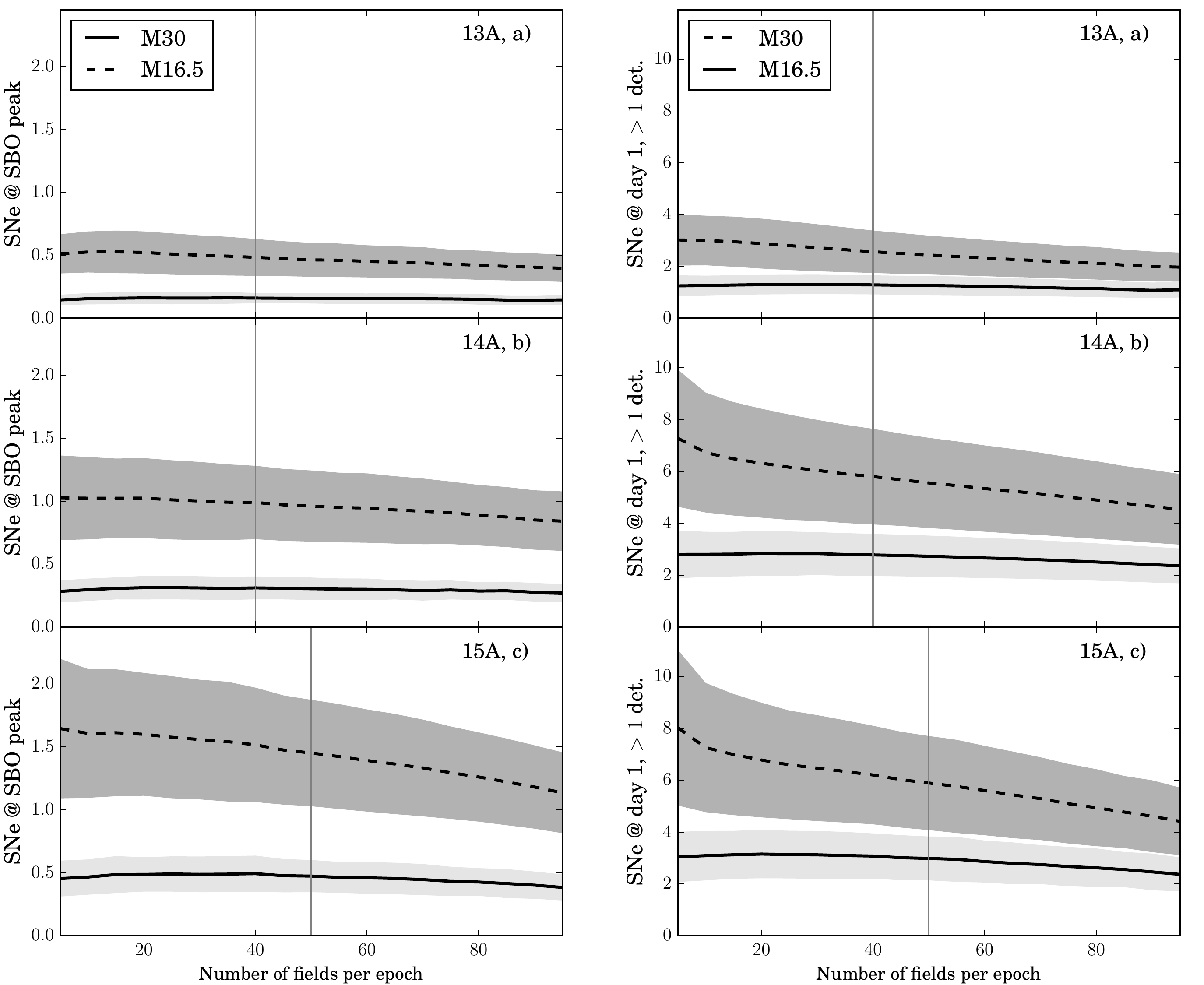} 
  \caption{{\bf a)} Simulated number of SBO optical peaks (see
    Figure~\ref{fig:SBO_LCs}) as a function of the number of fields
    observed per night for the 13A, 14A and 15A survey campaigns (see
    text for details). The shaded areas show the possible effect of
    the image difference process, ranging from the case where the
    noise associated with the reference image is negligible to the
    case where it has a noise level similar to the science image. {\bf
      b)} Same as a), but for the predicted number of SNe detectable
    at least twice during the first rest--frame day after shock
    emergence (see Figure~\ref{fig:day1_LCs}). The gray vertical lines
    correspond to the actual number of fields observed in the
    respective survey campaigns.}
\label{fig:simulations}
\end{figure*}

We found that using the ETC v5 instead of the ETC v6 led to an
overestimation of the number of SNe that would be detected during SBO
peak or twice during the first rest--frame day after shock emergence
of approximately 2.5 in both cases. Using the relation between airmass
and extinction from \citet{1983MNRAS.204..347S} to derive an excess
extinction at a given airmass led to an overestimation of the number of SNe
detected at SBO peak or twice within the same rest--frame day after
shock emergence of approximately 1.8 and 1.4, respectively. Moreover,
the loss in depth due to the difference imaging process can lead to an
overestimation of up to approximately 1.8 in both quantities.

Using the ETC v6 and a realistic airmass evolution we show the
predicted number of SNe that would be detected at SBO optical peak or
at least twice within the first rest--frame day after shock emergence
as a function of the number of fields per night per epoch in
Figure~\ref{fig:simulations}. We considered 4 continuous nights with 4
epochs per night in the $u$ band, 5 continuous nights with 4 epochs
per field per night in the $g$ band, and 6 continuous nights with 5
epochs per field per night in the $g$ band, as we did in 13A, 14A and
15A, respectively. We also consider the two $M_{\rm ZAMS}$
distributions discussed in Section~\ref{sec:simulations}, M16.5 and
M30, which differ in their upper mass cutoff limit.

Figure~\ref{fig:simulations} shows the predicted numbers of detections
using the modified DECam ETC v6, for the M16.5 and M30 averaged
ensemble of models as a function of the number of fields observed per
night for the 13A, 14A and 15A survey campaigns, as described in
Section~\ref{sec:simulations}. We assume 4 epochs per field per night
in the $u$ band during 4 continuous nights for 13A, 4 epochs per field
per night in the $g$ band during 5 continuous nights for 14A or 5
epochs per field per night in the $g$ band during 6 continuous nights
for 15A. Shorter exposure times have smaller $m_{\rm 50}$, which
favours more nearby SNe that are easier to follow up, but which
requires a faster real--time pipeline. We found no strong deviations
as a function of the number of fields within the parameter range under
consideration, i.e. wide/shallow surveys result in a similar number of
events than narrow/deep surveys, with less than 20\% and 30\%
variations with respect to the mean for the M16.5 and M30
distributions, respectively. The actual observational campaigns were
optimized using the ETC v5, which resulted in maxima at larger number
of fields per epoch than the ETC v6. Wider/shallower strategies enable
more follow up capabilities, favouring larger numbers of fields per
epoch or shorter exposure times depending on the available resources.

\begin{figure*}[!ht] 
  \includegraphics[scale = 0.7]{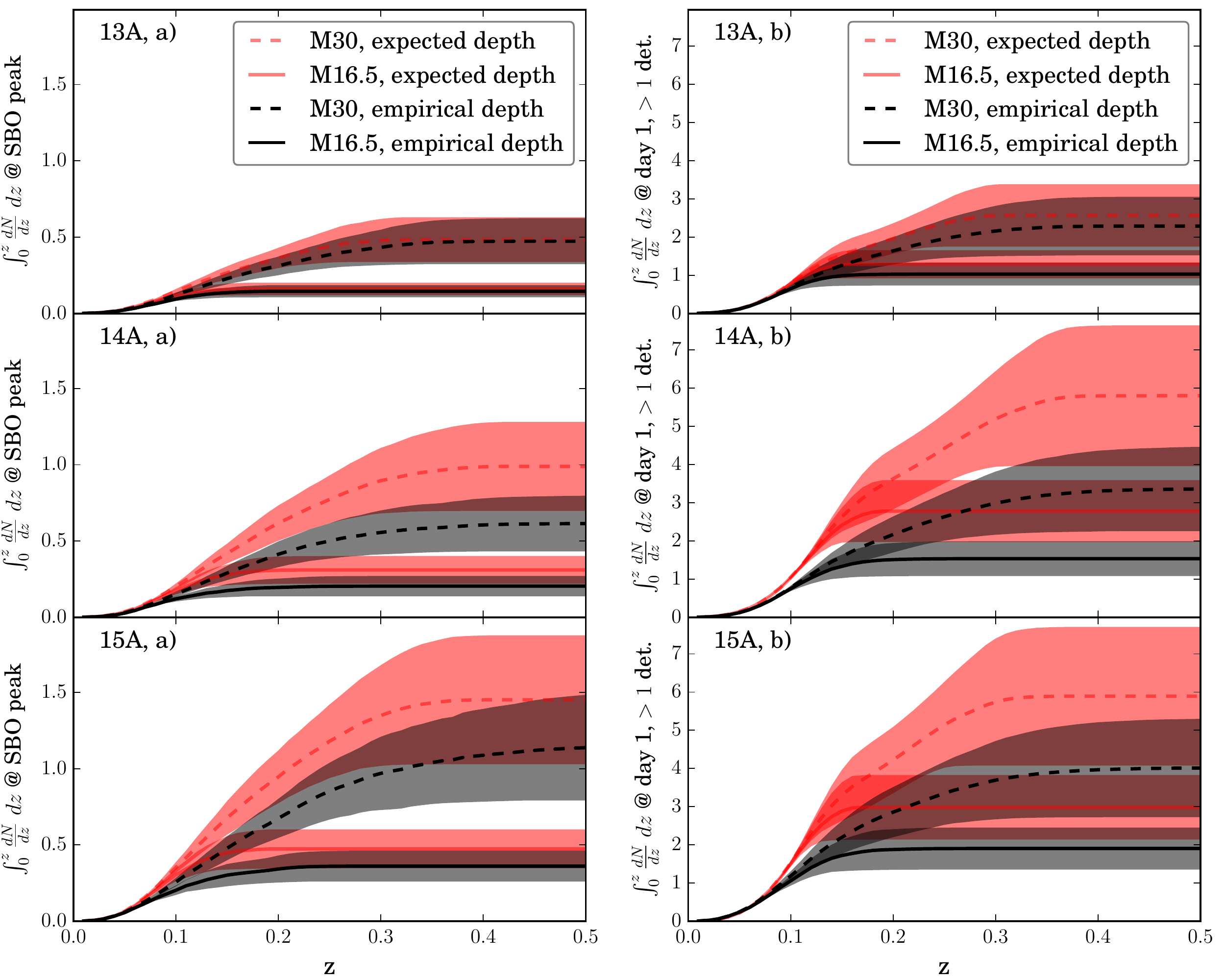} 
  \caption{Cumulative number of predicted SN detections during SBO
    optical peak {\bf (a)} or at least twice during the first
    rest--frame day after shock emergence {\bf (b)} as a function of
    redshift for the 13A, 14A and 15A survey campaigns. We compare the
    cumulative numbers in the cases when the limiting--magnitude is
    given by the ETC v6 with the expected FWHM, sky brightness and
    transparency and in the case where the empirical probability of
    detection, $P(m)$, is used (see Figure~\ref{fig:efficiencies}).
    The shaded areas show the possible effect of the image difference
    process as in Figure~\ref{fig:simulations}. }
\label{fig:cumulative}
\end{figure*}

In Figure~\ref{fig:cumulative} we compare the predicted number of SNe
computed using the empirical observational conditions (airmass, FWHM,
sky brightness and transparency) vs using the expected observational
conditions (FWHM, sky brightness and transparency), in order to
estimate the effect of cloud cover and bad seeing in the 13A, 14A and
15A campaigns. From this we conclude that in the 13A campaign the
atmospheric conditions led to an overestimation factor of
approximately 1.1 for the number of SNe that would have been
discovered at SBO optical peak and 1.2 for those discovered twice
within the first day after shock emergence. In the 14A campaign the
overestimation factors were approximately 1.4 and 1.6, respectively,
and in the 15A campaign, approximately 1.7 and 1.6, respectively.

In total, we estimate that the expected limiting--magnitudes (using
the unmodified ETCv5, the relation between extinction and airmass from
\citet{1983MNRAS.204..347S}, expected atmospheric conditions and
negligible reference image noise) led to a SBO number overestimation
factor of up to 14x compared to using the empirical
limiting--magnitudes.

\begin{figure*}
  \includegraphics[scale = 0.59]{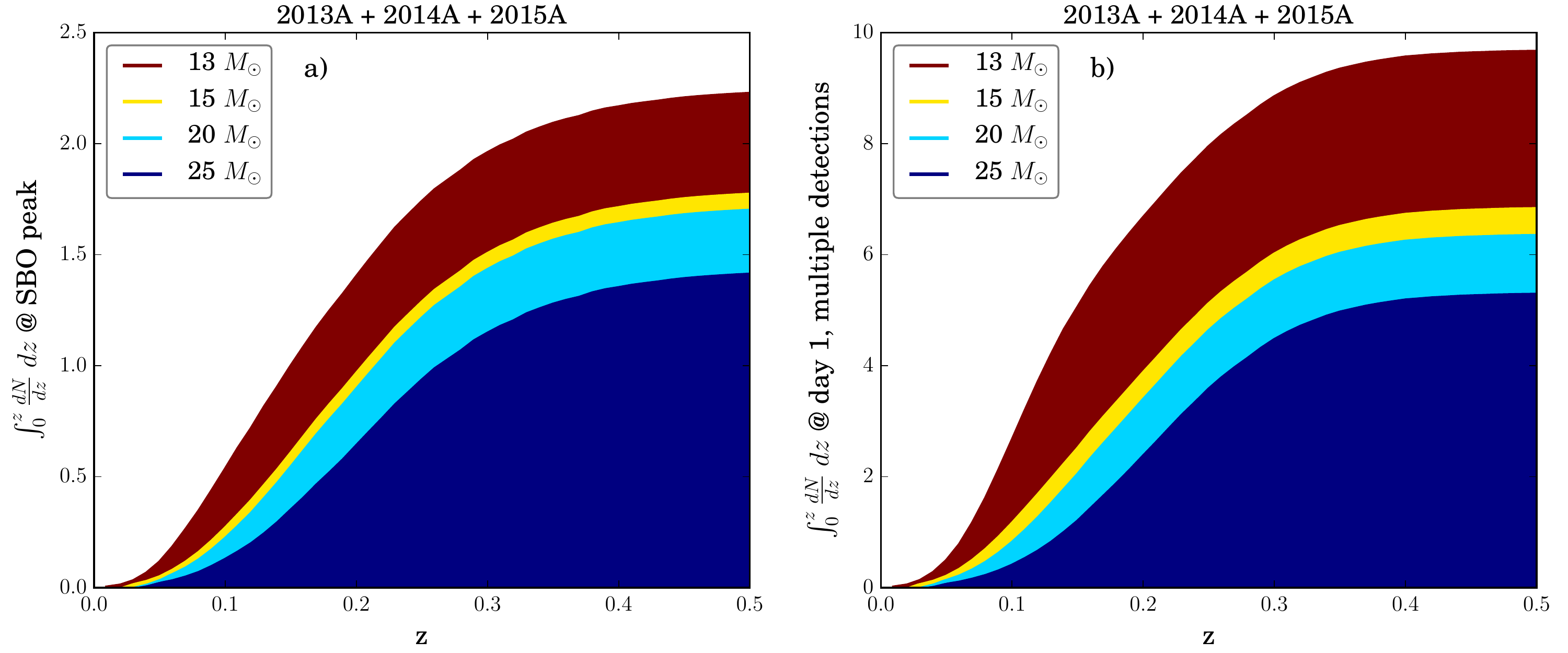} 
  \caption{Cumulative number of predicted SN detections during SBO
    optical peak {\bf (a)} or at least twice during the first
    rest--frame day after shock emergence {\bf (b)} as a function of
    redshift for the combined 13A, 14A and 15A survey campaigns
    assuming the M30 distribution and the \citet{2011ApJS..193...20T}
    models. We show the cumulative number of SNe obtained with the
    empirical probability of detection $P(m)$ and a perfect
    subtraction (the sum of the black thick lines in
    Figure~\ref{fig:cumulative}), but separating the contributions of
    each IMF--weighted explosion model as explained in
    Section~\ref{sec:simulations}.}
\label{fig:cumulative-models}
\end{figure*}

\subsection{Progenitor mass distribution}

The $M_{\rm ZAMS}$ distribution of the SNe found at SBO optical peak
or twice within the first rest--frame day after shock emergence should
differ considerably from the original SN II progenitor IMF
distribution. More massive progenitor scenarios that become observed
SNe II appear to have larger radii and more energetic explosions
\citep{2003ApJ...582..905H, 2016MNRAS.460..742M}, which would make
them brighter after shock emergence. This should bias our discoveries
towards larger $M_{\rm ZAMS}$ values in the M30 distribution. This can
be seen in the Figure~\ref{fig:cumulative-models}, where we combine
the 13A, 14A and 15A campaigns and separate the predicted cumulative
number of detections by associated progenitor model $M_{\rm ZAMS}$ for
the M30 distribution (we lack enough model resolution for the M16.5
distribution) and the \citet{2011ApJS..193...20T} models. In the M30
distribution, about 70--80\% of the expected SBO optical peak
detections should come from the weighted models with $M_{\rm ZAMS}$ of
20 or 25 $M_\odot$, and about 60--70\% in the case of the double
detections at day 1 after shock emergence for each survey
campaign. This means that the search for SBO or very young SNe will be
strongly biased towards the more massive SN II explosions.

\subsection{Parameter sensitivity}

The number of predicted events is sensitive to the upper mass limit of
the Type II supernova (SN II) initial mass function (IMF)
distribution, with a larger upper mass limit leading to more
detections as they are generally easier to detect than lower mass,
generally smaller radii explosions. As shown in
Figure~\ref{fig:cumulative-models}, an upper mass limit of 30
$M_\odot$ would lead to about four times more SBO detections than an
upper mass limit of 16.5 $M_\odot$, and about twice the number of SNe
younger than one rest--frame day after shock emergence with at least
two detections. We also tried a steeper IMF, with a $\gamma$ of 2.3
instead of 1.3, and we expect approximately 50\% fewer SBO detections
for the M30 distribution.

We have also explored the sensitivity to explosion energy. More
energetic models have brighter, narrower SBO optical peaks, and have
significantly brighter shock cooling optical light curves. The 25
$M_\odot$ $M_{\rm ZAMS}$ model from \citet{2011ApJS..193...20T}, which
is the dominant model in the M30 distribution, would produce 30\%
fewer or more optical SBO peak detections if instead of the 3 foe
model we had used the 1 and 9 foe models, respectively. However, it
would produce 2.7 and 2.6 times fewer or more SNe younger than one day
after shock emergence than the 3 foe model, respectively. Thus, the
explosion energy is not a very important parameter for the predicted
number of SBO detections, but it is a relevant parameter for the
number of very young SNe detected in their shock cooling phase.

We have also explored analytic models for the SBO
\citep{2010ApJ...725..904N} and early days of evolution of SNe II
\citep{2010ApJ...725..904N, 2011ApJ...728...63R} and have found that
they are effectively a two parameter family in the optical, where the
two parameters are the progenitor radius and the ratio between
explosion energy and progenitor mass. Larger radii, but also larger
energy to mass ratio explosions produce brighter/slower light curves
in these models. We have found that the models from
\citet{2010ApJ...725..904N} are generally in better agreement with
those from \citet{2011ApJS..193...20T} during the rise to maximum
light, but there are significant differences among all these models
which suggest that the physical interpretation of early SN light
curves should be treated with caution.

In addition to the model uncertainties, it is not clear how the SBO
optical peak should be observationally defined. We define the end of
the SBO optical peak for the \citet{2011ApJS..193...20T} models as the
time when 90\% of the magnitude change between the SBO peak and
subsequent minimum (maximum in magnitudes) has occured, which is shown
with thick lines in Figure~\ref{fig:SBO_LCs}. When trying a more
conservative definition, as the time when only 50\% of the previous
magnitude change has occurred, we found a decrease of only 1\% in the
number of predicted SBOs. If we define the SBO optical peak end as the
time when only 10\% of the previous magnitude change has occured we
see a decrease in the number of predicted SBOs of only 40\%. This low
sensitivity to the definition of the end of SBO optical peak means
that most predicted detections are expected to occur only once and
very close to the SBO optical peak maximum (minimum in magnitudes). In
fact, the number of SBO optical peak double detections is expected to
be very small, only 18\% of the SBO detections in the 90\% magnitude
difference definition, 3\% of the SBO detections in the 50\% magnitude
change definition and none in our simulations for the 10\% magnitude
difference definition.

\subsection{Constraints on the SBO optical peak} \label{sec:SBOlim}

\begin{figure}[!ht] 
  \includegraphics[scale = 0.4]{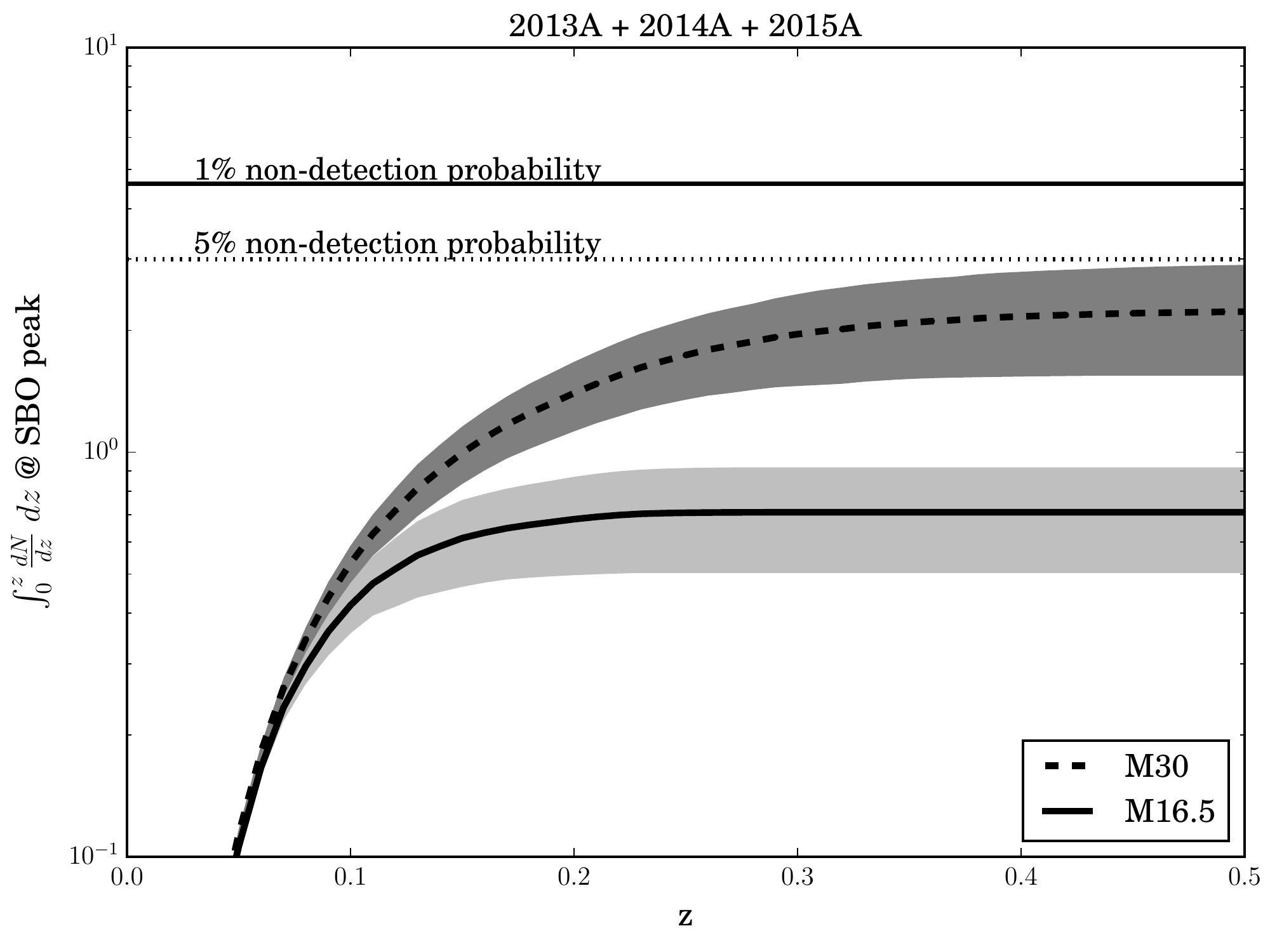} 
  \caption{Predicted cumulative number of SNe that should have been
    detected during SBO optical peak in the combined 13A, 14A and 15A
    campaigns assuming the M16.5 or M30 distributions described in
    Section~\ref{sec:simulations} for the selected models from
    \citet{2011ApJS..193...20T}, assuming the empirical detection
    probabilities, $P(m)$, and including the effect of the difference
    imaging as in previous figures. The horizontal dotted gray line
    indicates the expected number of events from which there is less
    than a 5\% chance of having no detections. }
\label{fig:modelconstraints}
\end{figure}

In Figure~\ref{fig:modelconstraints} we plot the cumulative
distribution of expected SBO detections as a function of redshift
using the \citet{2011ApJS..193...20T} models with the M16.5 and M30
distributions for the combined 13A, 14A and 15A campaigns. A dotted
horizontal gray line indicates the number of predicted events from
which a non--detection in the three survey campaigns has a probability
of less than 5\%. We conclude that although the search for SBO optical
peaks should be sensitive to the highest mass limit of the SN II IMF
distribution, the shallower than expected depth of the survey implies
that we cannot rule out either the M16.5 or M30 distributions for this
family of models.

\begin{figure}[!ht] 
  \includegraphics[scale = 0.4]{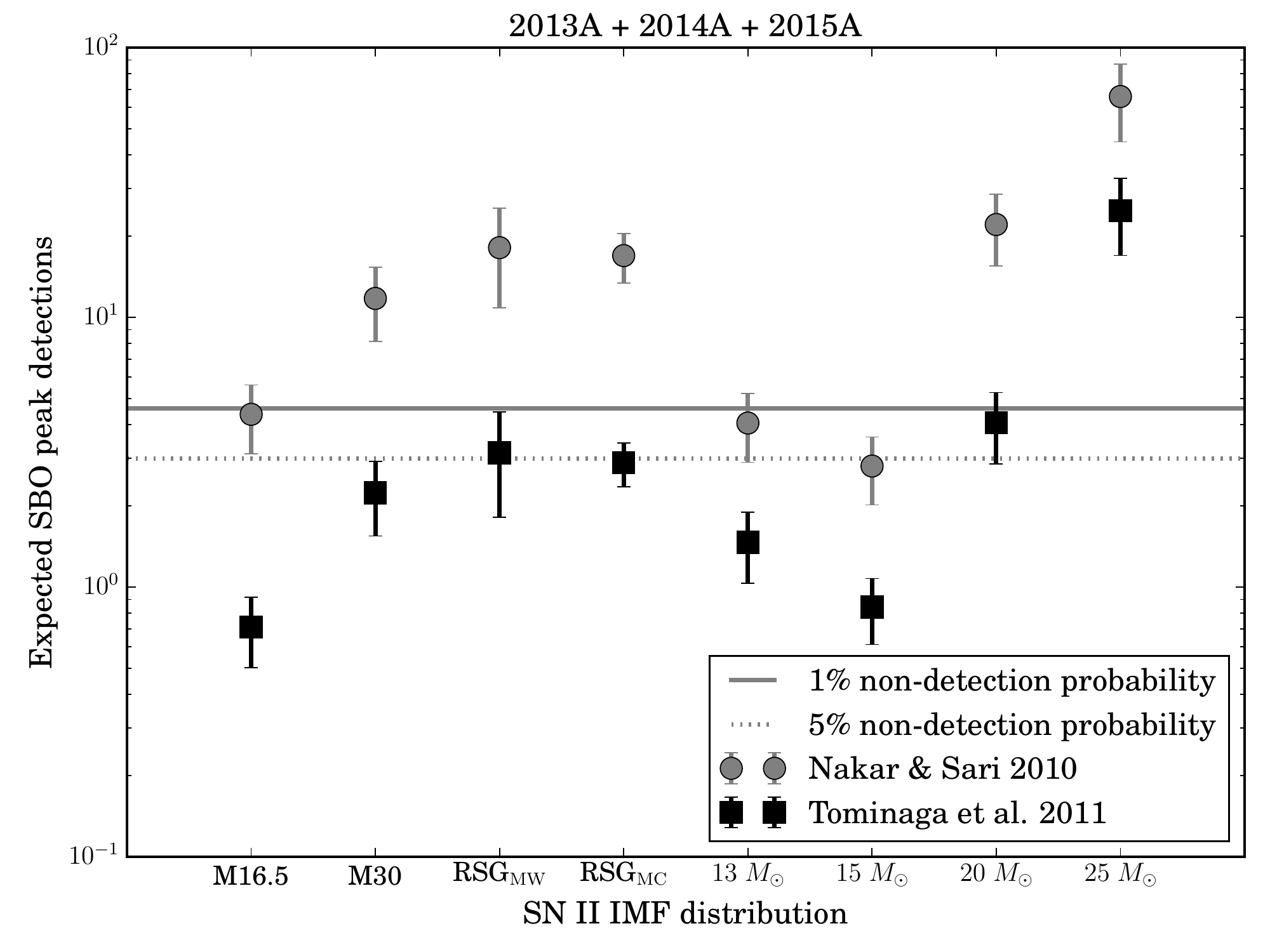} 
  \caption{Total number of expected SBO peak detections in the
    combined 13A, 14A and 15A campaigns assuming the empirical
    detections probabilities, $P(m)$, different SN II IMF
    distributions and the models from \citet{2011ApJS..193...20T} and
    \citet{2010ApJ...725..904N}. The vertical lines range from the
    case where the difference imaging introduces a negligible amount
    of additional noise to the case where it increases the noise by
    $\sqrt{2}$, as done in previous sections. We assume the
    Salpeter--like M16.5 and M30 distributions, as well as IMF
    distributions representative of observed RSG stars in the Milky
    Way and Magellanic clouds (see Section~\ref{sec:simulations}), and
    show for reference the (non--physical) case that all SBOs resemble
    those produced by each of the selected models of a given mass
    (using weights of 0.524 for the given model and zero for the
    rest).  The horizontal dotted and continuous gray lines indicate
    the expected number of events from which there is less than 5\%
    and 1\% probability of having no detections, respectively. We can
    marginally reject the brighter and longer--lived
    \citet{2010ApJ...725..904N} models under reasonable IMF
    distributions. We cannot reject the \citet{2011ApJS..193...20T}
    models under any reasonable IMF distribution, although we
    marginally reject that most SBOs resemble those produced by the
    more massive 20 and 25 $M_\odot$ models. See Table~\ref{tab:probs}
    for corresponding non--detection probabilities.}
\label{fig:allmodelconstraints}
\end{figure}

\begin{table}[h!]
  \centering
  \begin{tabular}{c | c c | c c } 
    \hline
    \hline
    SBO models & \multicolumn{2}{c}{Tominaga et al. 2011} & \multicolumn{2}{c}{Nakar \& Sari 2010} \\
    IMF dist. & min & max & min & max \\
    \hline 
    M16.5 & 0.40 & 0.60 & {\bf 0.003} & {\bf 0.04}\\
    M30 & {\bf 0.05} & 0.21 & $\bf 10^{-7}$ & $\bf 10^{-4}$ \\
    RSG$_{\rm MW}$ & {\bf 0.01} & 0.16 & $\bf 10^{-11}$ & $\bf 10^{-5}$ \\
    RSG$_{\rm MC}$ & {\bf 0.03} & 0.09 & $\bf 10^{-9}$ & $\bf 10^{-6}$ \\
    13 $M_\odot$ & 0.15 & 0.36 & {\bf 0.01} & 0.06 \\
    15 $M_\odot$ & 0.34 & 0.54 & {\bf 0.03} & 0.14 \\
    20 $M_\odot$ & {\bf 0.01} & {\bf 0.06} & $\bf 10^{-12}$ & $\bf 10^{-7}$ \\
    25 $M_\odot$ & $\bf 10^{-14}$ & $\bf 10^{-8}$ & $\bf 10^{-36}$ & $\bf 10^{-19}$ \\
    \hline
  \end{tabular}
  \caption{SBO non--detection probabilities for the combined 13A, 14A
    and 15A HITS campaigns assuming the SBO models from
    \citet{2011ApJS..193...20T} and \citet{2010ApJ...725..904N} and
    different SN II IMF distributions (see
    Figure~\ref{fig:allmodelconstraints}) with an optimistic image
    subtraction depth (negligible noise in the referece) and a
    conservative image subtraction depth (0.38 mag difference).}
    \label{tab:probs}
\end{table}

We then compare the previous results to those obtained using the
brighter and longer--lived \citet{2010ApJ...725..904N} models and
considering different model weigths which resembles observed RSG stars
discussed in Section~\ref{sec:simulations}. This is seen in
Figure~\ref{fig:allmodelconstraints}, where we show the total number
of events that should have been detected with different assumptions
about the SBO models, quality of the image difference and IMF
weighting scheme. The corresponding non--detection probabilities are
shown in Table~\ref{tab:probs}, where we have used that the
probability of having $k$ detections in a Poisson distribution with
mean $\lambda$ is $\lambda^k {\rm e}^{-\lambda} / k!$, which for $k=0$
is ${\rm e}^{-\lambda}$. The brighter and longer--lived models from
\citet{2010ApJ...725..904N} models can be marginally rejected under
all reasonable IMF distributions, specially the IMF distributions
which favour larger progenitor masses. We cannot reject the
\citet{2011ApJS..193...20T} models under any reasonable IMF
distribution, but we marginally reject that most SBOs resemble those
produced by their more massive 20 and 25 $M_\odot$ at ZAMS models
shown in Figures~\ref{fig:SBO_LCs} and \ref{fig:day1_LCs}.

\subsection{Supernova candidates}

Although no good SBO candidates were detected, several very young SN
candidates were detected in the three HiTS campaigns, a few of them
apparently less than a day after shock emergence, and more than a
hundred SN candidates of all ages in total, whose positions were made
available to the community \citep{2014ATel.5949....1F,
  2014ATel.5956....1F, 2015ATel.7099....1F, 2015ATel.7108....1F,
  2015ATel.7115....1F, 2015ATel.7122....1F, 2015ATel.7131....1F,
  2015ATel.7132....1F, 2015ATel.7146....1F, 2015ATel.7148....1F,
  2015ATel.7149....1F, 2015ATel.7221....1F, 2015ATel.7224....1F,
  2015ATel.7289....1F, 2015ATel.7290....1F}. We could
spectroscopically classify about a dozen of these \citep[see
  e.g.][]{2014ATel.5957....1W, 2014ATel.5970....1W,
  2014ATel.6014....1A, 2015ATel.7162....1A, 2015ATel.7164....1A,
  2015ATel.7335....1A, 2015ATel.7291....1F}, confirming their SN
nature in all cases. We have also used the models from
\citet{2011ApJS..193...20T} to study the number of SNe II expected to
be detected twice during their rise (as our template images were taken
at the beginning of each run we should only detect rising SNe) and
found that we expect about 80 and 130 SNe II for the M16.5 and M30
distributions, respectively. Because these numbers are derived using a
supernova efficiency computed from actual observations
\citep{2015ApJ...813...93S}, this suggests that the total number of
events we detect is roughly consistent with the observed rate of SNe
within the relevant redshift range. Considering a contribution from
other SN types (which will be the subject of future publications) and
model uncertainties this also suggests that the presence of other
types of transients is not required to explain the number of detected
events in our sample, and may even suggest a preference for the M16.5
distribution.  We will present a catalogue of calibrated light curves
which will include follow up resources other than DECam in a future
publication, including host galaxy redshifts and post--SN images for
the most interesting candidates. Interestingly, the SN candidates
apparently younger than one day after shock emergence appear to show
initially faster than expected rise times
\citep[c.f.][]{2015MNRAS.451.2212G, 2016arXiv160103261T,
  2016ApJ...820...23G}, which suggests that the evolution of SN II
light curves during the first days may not be explained by standard
assumptions about their circumstellar material.

\section{SUMMARY} \label{sec:summary}

The first results of the High cadence Transient Survey (HiTS) search
for supernova (SN) shock breakouts (SBOs) were presented. With the
current calibration scheme and data analysis pipeline we see no clear
evidence for red supergiant SBO optical early--time optical peaks in
light curves resembling SNe II (with a SNR $> 5$). Based on a joint
analysis of the three observational campaigns with our empirically
derived limiting--magnitudes we conclude that ensembles of explosion
models assuming a Salpeter--like initial mass function with an upper
mass limit of either 16.5 or 30 $M_\odot$ (M16.5 or M30 distributions,
respectively) cannot be excluded for the \citet{2011ApJS..193...20T}
models, but we can be marginally excluded for the
\citet{2010ApJ...725..904N} models (see Table~\ref{tab:probs}). This
result should be taken with caution given all the uncertainties
associated to the distribution of SN II progenitor properties.

HiTS run in the optical using the Dark Energy Camera (DECam) during
the 13A, 14A and 15A survey campaigns. The survey strategy could be
described as several contiguous nights (4 nights in 13A, 5 nights in
14A and 6 nights in 15A) of high cadence (2 hr in 13A and 14A and 1.6
hr in 15A), monochromatic ($u$ band in 13A and $g$ band in 14A and
15A), untargeted, varying airmass observations towards a large area of
the sky (120 deg$^2$ in 13A and 14A and 150 deg$^2$ in 15A) with
single epoch depths between 23 and 24.5 magnitudes.

In March of 2013 a pilot phase of the survey was performed, with data
being analyzed after the observation run had finished. In March of
2014 we performed the data processing and candidate filtering in
real--time, the first real--time analysis of DECam data to our
knowledge, although with very simple visualization tools which delayed
our reaction capability by a few hours. In February 2015 we achieved
the full data analysis, candidate filtering and visualization process
in real--time thanks to significant improvements in our visualization
tools, which highlights the importance of fast visualization for
real--time surveys. We processed more than $10^{12}$ pixels in a
stream of 40 Mbps, which after processing resulted in a stream of
about 3 new candidates per minute, 5 to 6 minutes after every
exposure. As a result, more than 120 SN candidates were detected in
total in real--time.

We computed empirical 50\% completeness magnitudes analyzing deep
stacked DECam images in relation to the individual epochs of the
survey. The depth of the survey varied within each night typically by
more than one magnitude as the observations had to be performed at
varying airmasses in order to achieve the required cadence during the
full night. We compared these values with the predictions of the
public exposure time calculators (ETCs) versions 5 and 6 (v5 and v6,
respectively), which we modified to include the effect of airmass. We
validated our modified ETCs studying the relation between these
limiting--magnitudes with the observed FWHM and airmasses. The elusive
SBOs may have been detected if the actual survey depth had matched the
initial limiting--magnitude estimations. However, the survey depth was
overestimated due to several factors: an overly optimistic ETC
available at the time (ETC v5), a stronger than expected airmass
effect, worse than planned observing conditions and errors introduced
by the image difference process.

During survey design we used two figures of merit to determine the
quality of an observational strategy, both defined in
Section~\ref{sec:simulations}: 1) the number of SBO optical peak
detections and 2) the number of SNe detected at least twice during the
first rest--frame day after shock emergence. We have shown that a SBO
optical peak detection is much harder to obtain than a double
detection of the SN within the first rest--frame day. An obvious
consequence is that an example of 2) does not mean that the SBO
optical peak should have been seen in the data, which may explain why
we have not seen optical SBO peaks in some datasets \citep[see
  e.g.][]{2016ApJ...818....3K}.

Using the empirical limiting--magnitudes with the models described in
this analysis \citep[from][]{2011ApJS..193...20T, 2010ApJ...725..904N}
we evaluated these two figures of merit and found that the number of
predicted events was most sensitive to the upper mass limit of the SN
II initial mass function (IMF) distribution. An upper mass limit of 30
$M_\odot$ would lead to about four times more SBO detections than an
upper mass limit of 16.5 $M_\odot$, and about twice the number of SNe
younger than one day after shock emergence with at least two
detections. The number of SNe detected twice during the first day
after shock emergence appears to be even more sensitive to the
explosion energy, varying by as much as the energy variation factor in
the models tested.

An important consequence of marginally favouring the relatively dimmer
and shorter--lived SBO optical peaks from the
\citet{2011ApJS..193...20T} models is that with our typical cadences a
real--time detection of a SBO will be unlikely to happen fast enough
to react and observe it with other instruments. We expect more than
82\% of our SBO detections to have only one detection before the end
of the SBO optical peak with these models, thus we rely on the
subsequent early rising SN light for their online identification. In
fact, for the shock cooling SN light curve to rise to at least half a
magnitude below the SBO optical peak maximum it takes typically more
than half a day according to the more realistic
\citet{2011ApJS..193...20T} models (see Figure~\ref{fig:day1_LCs}),
making the identification of SNe within a few hours of shock emergence
incredibly challenging. This highlights the importance of having
continuous high cadence observations during several nights followed by
low cadence follow up observations in order to aid with the candidate
selection and SBO identification via post--processing of the high
cadence phase data.

\section{IMPLICATIONS AND FUTURE WORK} \label{sec:discussion}

Apart from the previously described SBO model constraints, an
important contribution from this survey will be the detection of SN
candidates younger than one day after shock emergence. A preliminary
analysis show that they have a very fast initial rise in their light
curve inconsistent with the model light curves used in this
analysis. They seem more consistent with shocks breaking into high
density circumstellar material (CSM) in RSG stars
\citep{2015ApJ...804...28G, 2015MNRAS.451.2212G, 2016ApJ...818....3K,
  2016arXiv160103261T, 2016ApJ...820...23G}. In order to compare their
light curves to existing models we are compiling host galaxy redshifts
and post--SN explosion images for more precise and accurate absolute
calibrations.

If the shock--CSM interaction in normal RSG stars significantly
affects the SBO properties, which could be the case for the high
density CSM shock breakouts, the evolution of SN II light curves
during the first rest--frame day after shock emergence could be very
different to that suggested by the models considered in this
work. Thus, the detection of SNe during the first days after shock
emergence could be a tool to constrain the properties of RSGs and
their CSM. Given the discrepancy in derived mass loss rates between
early and late times implied by these recent works, this could be a
clue about the wind structure in RSGs \citep{2014Natur.512..282M} or
the latest stages of nuclear burning before explosion in these stars
\citep{2014MNRAS.445.2492M}. These factors and our non--detection of
red supergiant SBOs suggest that HiTS should switch to a lower
cadence, multiwavelength survey mode.

If SBOs can be detected in a systematic fashion, they could provide an
alternative probe for the upper mass limit of the SN II IMF
distribution than pre--explosion progenitor detections
\citep{2009MNRAS.395.1409S}. This is because the UV radiation during
the SBO phase is expected to sublimate most of the CSM dust which
could be obscuring the SN II progenitors in pre--explosion images.
Some evidence for dust column density changes after SBO exist for SN
2012aw \citep{2012ApJ...759L..13F, 2012ApJ...756..131V}, which appears
to be on the high end of the progenitor mass distribution (although
see \citealt{2012ApJ...759...20K}).

The high cadence strategy will be very important for future SBO
surveys, since it is difficult to confirm a SBO detection without a
previous non--detection and a subsequent drop in the light curve with
a timescale comparable to the SBO optical peak. In a multichromatic
survey it may be possible to differentiate SBOs from early SN light
curves based on their colors, but this requires either simultaneous
multiwavelength observations or high cadence filter changes, which may
be expensive for large etendue telescopes. In the case of LSST,
intranight multicolor observations will be limited because of the
limited number of allowed filter changes during its lifetime, so most
SBO detections will most likely rely on high cadence monochromatic
observations in the deep--drilling fields to be defined.

We have scaled our simulations to LSST's larger FoV, larger mirror
area, smaller pixel size and shorter overhead times and found that
assuming 30 sec exposure times we could observe 170 LSST fields per
night with a cadence of 1.6 hours and would be able to find 13 times
more SBO and very young SNe than in our 15A strategy assuming 6
continuous nights of observations. Given that in our simulations 30
sec LSST exposures would produce similar limiting--magnitudes than our
87 sec DECam exposures (a simple scaling suggests that 30 sec $g$ band
LSST observations would be 0.1 mag deeper than 87 sec $g$ band DECam
observations), this can be approximately explained by a combination of
an increased number of fields that can be observed up to a
limiting--magnitude during one night (3.4 x larger) and LSST's larger
FoV (3.2 x), and is only slightly more than what would be obtained
scaling the two telescope's etendues. We did a similar exercise for
the Hyper Suprime--Cam instrument and we obtained a similar result,
i.e. that the predicted number of SBO and young SNe should
approximately scale with etendue.

Continuous observations with large etendue telescopes from space or
from polar regions of the earth could provide high cadence
observations without the large airmass and limiting--magnitude
variations experienced by this survey. In fact, the recent detection
of a red supergiant SBO candidate with the large etendue, space--based
\emph{Kepler} observatory \citep{2016ApJ...820...23G} combines both
the continuous high cadence and large etendue required for this
purpose. Alternatively, moderately large etendue, space based X--ray
or UV observatories may be better tools to look for SBOs in BSG and
RSG stars, but with the \emph{GALEX} mission now completed we will
need to wait for future space missions for this to happen
\citep[e.g.][]{2002SPIE.4497..115F}. Interestingly, gravitational
waves \citep{2016PhRvL.116f1102A} or neutrino detections
\citep{1987PhRvL..58.1490H, 2013ApJ...778...81K} may be complementary
methods for obtaining earth based high cadence observations of SBOs if
candidates with a signature consistent with core collapse SNe in
nearby galaxies within the detectors localization errors are targeted,
but this will require either very wide field--of--view telescopes or
arrays of robotic telescopes observing several galaxies simultaneously
and able to reach the necessary absolute magnitudes shown in
Figure~\ref{fig:SBO_LCs}.

Finally, it should be noted that the current observation and data
analysis strategy followed by HiTS allowed us to study the optical
Universe with an unprecedented combination of total observed volume,
high cadence for several continuous nights and real--time data
reduction including visualization, emphasizing the importance of
interdisciplinary collaborations for astronomy. These observations are
currently being used as a verification dataset for the LSST software
stack \footnote{http://dm.lsst.org/} and will likely become a legacy
dataset for different scientific purposes.

\section{ACKNOWLEDGMENTS}

We kindly thank N.~Tominaga for providing detailed explosion models
used in this analysis. We also thank C.~Aragon, M.~Graham, K. Maeda,
R.~Mu\~noz, E.~Nakar, P.~Protopapas and F.~Valdes for useful
discussions, as well as the PESSTO collaboration for providing
spectroscopic confirmations. F.F., J.S.M and S.G. acknowledge support
from Basal Project PFB--03. F.F., M.H., G.P., H.K., F.O., L.G., S.G.,
P.E. acknowledge support from the Ministry of Economy, Development,
and Tourism's Millennium Science Initiative through grant IC120009,
awarded to The Millennium Institute of Astrophysics (MAS).
F.F. acknowledges support from Conicyt through the Fondecyt Initiation
into Research project No. 11130228.  F.O., S.G., P.H. and
G.C. acknowledge support from FONDECYT postdoctoral grants 3140326,
3130680 and 3150460 and 3140566, respectively. F.F., J.C.M., P.H.,
G.C., P.E. acknowledge support from Conicyt through the Programme of
International Cooperation project DPI20140090. G.P. acknowledges
support by the Proyecto Regular FONDECYT 1140352. We acknowledge
support from Conicyt through the infrastructure Quimal project
No. 140003. Powered@NLHPC: this research was partially supported by
the supercomputing infrastructure of the NLHPC (ECM-02).  This project
used data obtained with the Dark Energy Camera (DECam), which was
constructed by the Dark Energy Survey (DES) collaboration. Funding for
the DES Projects has been provided by the U.S. Department of Energy,
the U.S. National Science Foundation, the Ministry of Science and
Education of Spain, the Science and Technology Facilities Council of
the United Kingdom, the Higher Education Funding Council for England,
the National Center for Supercomputing Applications at the University
of Illinois at Urbana--Champaign, the Kavli Institute of Cosmological
Physics at the University of Chicago, Center for Cosmology and
Astro--Particle Physics at the Ohio State University, the Mitchell
Institute for Fundamental Physics and Astronomy at Texas A\&M
University, Financiadora de Estudos e Projetos, Funda\c{c}\~ao Carlos
Chagas Filho de Amparo, Financiadora de Estudos e Projetos,
Funda\c{c}\~ao Carlos Chagas Filho de Amparo \`a Pesquisa do Estado do
Rio de Janeiro, Conselho Nacional de Desenvolvimento Cient\'ifico e
Tecnol\'ogico and the Minist\'erio da Ci\^encia, Tecnologia e
Inova\c{c}\~ao, the Deutsche Forschungsgemeinschaft and the
Collaborating Institutions in the Dark Energy Survey. The
Collaborating Institutions are Argonne National Laboratory, the
University of California at Santa Cruz, the University of Cambridge,
Centro de Investigaciones Energ\'eticas, Medioambientales y
Tecnol\'ogicas–Madrid, the University of Chicago, University College
London, the DES--Brazil Consortium, the University of Edinburgh, the
Eidgen\"ossische Technische Hochschule (ETH) Z\"urich, Fermi National
Accelerator Laboratory, the University of Illinois at
Urbana--Champaign, the Institut de Ci\`encies de l'Espai (IEEC/CSIC),
the Institut de F\'isica d'Altes Energies, Lawrence Berkeley National
Laboratory, the Ludwig--Maximilians Universit\"at M\"unchen and the
associated Excellence Cluster Universe, the University of Michigan,
the National Optical Astronomy Observatory, the University of
Nottingham, the Ohio State University, the University of Pennsylvania,
the University of Portsmouth, SLAC National Accelerator Laboratory,
Stanford University, the University of Sussex, and Texas A\&M.
University.

\appendix
\section{APPENDIX: HiTS fields}
\begin{deluxetable}{c c c c}
  \tabletypesize{\small}
  \tablewidth{0pt}
  \centering \tablecaption{Fields observed during the 2013A semester,
    mostly in the $u$ band. 4 epochs per night per field were observed during 4
    consecutive nights. The semester, field name, right ascension (RA)
    and declination (DEC) are shown.}  \tablehead{\colhead{Semester} &
    \colhead{Field name} & \colhead{RA} & \colhead{DEC} \\ & &
    \colhead{[hr]} & \colhead{[deg]} } \startdata 2013A & Blind13A\_01 & 10:36:32.960 & -29:11:29.00 \\
2013A & Blind13A\_02 & 10:43:03.600 & -27:05:42.80 \\
2013A & Blind13A\_03 & 10:38:35.450 & -25:00:03.20 \\
2013A & Blind13A\_04 & 10:44:42.390 & -22:54:12.40 \\
2013A & Blind13A\_05 & 10:55:12.630 & -22:54:15.10 \\
2013A & Blind13A\_06 & 10:49:16.180 & -24:59:59.10 \\
2013A & Blind13A\_07 & 10:53:56.110 & -27:05:44.60 \\
2013A & Blind13A\_08 & 10:47:38.500 & -29:11:28.00 \\
2013A & Blind13A\_09 & 10:58:43.680 & -29:11:28.20 \\
2013A & Blind13A\_10 & 11:04:48.510 & -27:05:43.70 \\
2013A & Blind13A\_11 & 10:59:57.150 & -24:59:58.30 \\
2013A & Blind13A\_12 & 11:05:43.380 & -22:54:13.40 \\
2013A & Blind13A\_13 & 11:16:13.750 & -22:54:14.60 \\
2013A & Blind13A\_14 & 11:10:37.970 & -24:59:58.60 \\
2013A & Blind13A\_15 & 11:15:41.070 & -27:05:43.30 \\
2013A & Blind13A\_16 & 11:09:49.250 & -29:11:27.40 \\
2013A & Blind13A\_17 & 11:20:54.390 & -29:11:27.90 \\
2013A & Blind13A\_18 & 11:26:33.450 & -27:05:42.80 \\
2013A & Blind13A\_19 & 11:21:19.040 & -24:59:57.40 \\
2013A & Blind13A\_20 & 11:26:44.600 & -22:54:13.20 \\
2013A & Blind13A\_21 & 11:37:14.920 & -22:54:13.70 \\
2013A & Blind13A\_22 & 11:31:59.860 & -24:59:57.70 \\
2013A & Blind13A\_23 & 11:37:26.010 & -27:05:42.50 \\
2013A & Blind13A\_24 & 11:31:59.970 & -29:11:26.50 \\
2013A & Blind13A\_25 & 11:43:05.120 & -29:11:27.40 \\
2013A & Blind13A\_26 & 11:48:18.410 & -27:05:41.90 \\
2013A & Blind13A\_27 & 11:42:40.870 & -24:59:56.50 \\
2013A & Blind13A\_28 & 11:47:45.780 & -22:54:12.00 \\
2013A & Blind13A\_29 & 11:58:16.120 & -22:54:12.70 \\
2013A & Blind13A\_30 & 11:53:21.710 & -24:59:56.40 \\
2013A & Blind13A\_31 & 11:59:10.950 & -27:05:41.50 \\
2013A & Blind13A\_32 & 11:54:10.700 & -29:11:25.30 \\
2013A & Blind13A\_33 & 12:05:15.840 & -29:11:26.10 \\
2013A & Blind13A\_34 & 12:10:03.360 & -27:05:41.00 \\
2013A & Blind13A\_35 & 12:04:02.680 & -24:59:55.10 \\
2013A & Blind13A\_36 & 12:08:46.960 & -22:54:10.60 \\
2013A & Blind13A\_37 & 12:19:17.350 & -22:54:10.70 \\
2013A & Blind13A\_38 & 12:14:43.550 & -24:59:55.20 \\
2013A & Blind13A\_39 & 12:20:55.930 & -27:05:40.30 \\
2013A & Blind13A\_40 & 12:16:21.420 & -29:11:23.80

  \enddata
\label{tab:2013Afields}
\end{deluxetable}

\begin{deluxetable}{c c c c c}
  \tabletypesize{\small}
  \tablewidth{0pt}
  \centering \tablecaption{Fields observed during the 2014A semester,
    mostly in the $g$ band. 4 epochs per night per field were observed
    during 5 consecutive nights. The semester, field name, right
    ascension (RA) and declination (DEC) are shown, as well as the
    2015A field name for those matching the positions of fields observed
    during the 2015A semester.}  \tablehead{\colhead{Semester} &
    \colhead{Field name} & \colhead{RA} & \colhead{DEC} &
    \colhead{2015A field name}\\ & & \colhead{[hr]} & \colhead{[deg]}}
  \startdata 2014A & Blind14A\_01 & 10:08:46.320 & -02:05:44.81 & Blind15A\_33 \\
2014A & Blind14A\_02 & 10:00:20.640 & -02:05:44.81 & Blind15A\_28 \\
2014A & Blind14A\_03 & 09:58:48.000 & +00:09:00.00 & Blind15A\_27 \\
2014A & Blind14A\_04 & 10:00:28.800 & +02:12:36.00 & Blind15A\_26 \\
2014A & Blind14A\_05 & 10:09:32.880 & +02:05:44.81 & Blind15A\_35 \\
2014A & Blind14A\_06 & 10:12:28.800 & +00:00:00.00 & Blind15A\_34 \\
2014A & Blind14A\_07 & 10:17:17.760 & -02:05:44.81 & Blind15A\_38 \\
2014A & Blind14A\_08 & 10:12:22.560 & -04:11:29.62 & Blind15A\_39 \\
2014A & Blind14A\_09 & 10:21:52.800 & -04:57:00.00 & Blind15A\_42 \\
2014A & Blind14A\_10 & 10:20:28.800 & -06:31:12.00 & Blind15A\_40 \\
2014A & Blind14A\_11 & 10:22:00.000 & -08:06:00.00 \\
2014A & Blind14A\_12 & 10:25:00.000 & -09:34:12.00 \\
2014A & Blind14A\_13 & 10:32:48.000 & -08:48:00.00 \\
2014A & Blind14A\_14 & 10:36:35.520 & -06:17:14.42 & Blind15A\_50 \\
2014A & Blind14A\_15 & 10:31:47.280 & -04:11:29.62 & Blind15A\_49 \\
2014A & Blind14A\_16 & 10:36:40.320 & -02:05:44.81 & Blind15A\_48 \\
2014A & Blind14A\_17 & 10:41:29.760 & -04:11:29.62 \\
2014A & Blind14A\_18 & 10:46:21.360 & -02:05:44.81 \\
2014A & Blind14A\_19 & 10:41:31.200 & +00:00:00.00 & Blind15A\_47 \\
2014A & Blind14A\_20 & 10:46:21.360 & +02:05:44.81 \\
2014A & Blind14A\_21 & 10:56:02.640 & +02:05:44.81 \\
2014A & Blind14A\_22 & 10:51:12.000 & +00:00:00.00 \\
2014A & Blind14A\_23 & 10:56:02.640 & -02:05:44.81 \\
2014A & Blind14A\_24 & 10:51:12.000 & -04:11:29.62 \\
2014A & Blind14A\_25 & 11:00:54.240 & -04:11:29.62 \\
2014A & Blind14A\_26 & 11:05:43.680 & -02:05:44.81 \\
2014A & Blind14A\_27 & 11:00:52.800 & +00:00:00.00 \\
2014A & Blind14A\_28 & 11:05:43.680 & +02:05:44.81 \\
2014A & Blind14A\_29 & 11:15:24.960 & +02:05:44.81 \\
2014A & Blind14A\_30 & 11:10:33.600 & +00:00:00.00 \\
2014A & Blind14A\_31 & 11:15:24.960 & -02:05:44.81 \\
2014A & Blind14A\_32 & 11:10:36.720 & -04:11:29.62 \\
2014A & Blind14A\_33 & 11:20:18.960 & -04:11:29.62 \\
2014A & Blind14A\_34 & 11:25:06.240 & -02:05:44.81 \\
2014A & Blind14A\_35 & 11:20:14.400 & +00:00:00.00 \\
2014A & Blind14A\_36 & 11:25:06.240 & +02:05:44.81 \\
2014A & Blind14A\_37 & 11:34:47.280 & +02:05:44.81 \\
2014A & Blind14A\_38 & 11:29:55.200 & +00:00:00.00 \\
2014A & Blind14A\_39 & 11:34:47.280 & -02:05:44.81 \\
2014A & Blind14A\_40 & 11:30:01.440 & -04:11:29.62
 \enddata
  \vspace{-0.6cm}
\label{tab:2014Afields}
\end{deluxetable}

\begin{deluxetable}{c c c c c} 
  \tabletypesize{\small}
  \tablewidth{0pt}
  \centering \tablecaption{Fields observed during the 2015A semester,
    mostly in the $g$ band. 5 epochs per night per field were observed
    during 6 consecutive nights followed by three non--consecutive
    half nights. The semester, field name, right ascension (RA) and
    declination (DEC) are shown, as well as the 2014A field name for
    those matching the positions of fields observed during the 2014A
    semester.}  \tablehead{\colhead{Semester} & \colhead{Field name} &
    \colhead{RA} & \colhead{DEC} & \colhead{2014A field name} \\ & &
    \colhead{[hr]} & \colhead{[deg]} } \startdata 2015A & Blind15A\_01 & 09:13:31.152 & -06:18:04.68 \\
2015A & Blind15A\_02 & 09:08:53.952 & -04:18:02.81 \\
2015A & Blind15A\_03 & 09:13:30.840 & -02:17:49.30 \\
2015A & Blind15A\_04 & 09:08:53.952 & -00:17:59.06 \\
2015A & Blind15A\_05 & 09:13:31.152 & +01:42:02.81 \\
2015A & Blind15A\_06 & 09:22:45.552 & +01:42:02.81 \\
2015A & Blind15A\_07 & 09:18:08.352 & -00:17:59.06 \\
2015A & Blind15A\_08 & 09:22:45.552 & -02:18:00.94 \\
2015A & Blind15A\_09 & 09:18:08.352 & -04:18:02.81 \\
2015A & Blind15A\_10 & 09:22:45.552 & -06:18:04.68 \\
2015A & Blind15A\_11 & 09:31:59.952 & -06:18:04.68 \\
2015A & Blind15A\_12 & 09:27:22.752 & -04:18:02.81 \\
2015A & Blind15A\_13 & 09:31:59.952 & -02:18:00.94 \\
2015A & Blind15A\_14 & 09:27:22.752 & -00:17:59.06 \\
2015A & Blind15A\_15 & 09:31:59.952 & +01:42:02.81 \\
2015A & Blind15A\_16 & 09:41:14.352 & +01:42:02.81 \\
2015A & Blind15A\_17 & 09:36:37.152 & -00:17:59.06 \\
2015A & Blind15A\_18 & 09:41:14.352 & -02:18:00.94 \\
2015A & Blind15A\_19 & 09:36:37.152 & -04:18:02.81 \\
2015A & Blind15A\_20 & 09:41:14.352 & -06:18:04.68 \\
2015A & Blind15A\_21 & 09:50:28.752 & -06:18:04.68 \\
2015A & Blind15A\_22 & 09:45:51.552 & -04:18:02.81 \\
2015A & Blind15A\_23 & 09:50:28.752 & -02:18:00.94 \\
2015A & Blind15A\_24 & 09:49:33.312 & -00:04:07.46 \\
2015A & Blind15A\_25 & 09:50:28.752 & +02:09:46.01 \\
2015A & Blind15A\_26 & 10:00:28.800 & +02:12:36.00 & Blind14A\_04 \\
2015A & Blind15A\_27 & 09:58:48.000 & +00:09:00.00 & Blind14A\_03 \\
2015A & Blind15A\_28 & 10:00:20.760 & -02:05:44.81 & Blind14A\_02 \\
2015A & Blind15A\_29 & 09:55:43.560 & -04:05:46.68 \\
2015A & Blind15A\_30 & 10:00:20.760 & -06:05:48.55 \\
2015A & Blind15A\_31 & 10:09:35.160 & -06:05:48.55 \\
2015A & Blind15A\_32 & 10:04:02.520 & -04:05:46.68 \\
2015A & Blind15A\_33 & 10:08:46.394 & -02:05:44.81 & Blind14A\_01 \\
2015A & Blind15A\_34 & 10:12:28.800 & +00:00:00.00 & Blind14A\_06 \\
2015A & Blind15A\_35 & 10:09:32.887 & +02:05:44.81 & Blind14A\_05 \\
2015A & Blind15A\_36 & 10:18:47.287 & +02:05:44.81 \\
2015A & Blind15A\_37 & 10:21:33.607 & +00:05:42.94 \\
2015A & Blind15A\_38 & 10:17:17.839 & -02:05:44.81 & Blind14A\_07 \\
2015A & Blind15A\_39 & 10:12:22.570 & -04:11:29.62 & Blind14A\_08 \\
2015A & Blind15A\_40 & 10:20:28.800 & -06:31:12.00 & Blind14A\_10 \\
2015A & Blind15A\_41 & 10:28:47.760 & -06:31:12.00 \\
2015A & Blind15A\_42 & 10:21:52.800 & -04:57:00.00 & Blind14A\_09 \\
2015A & Blind15A\_43 & 10:27:25.440 & -02:08:27.52 \\
2015A & Blind15A\_44 & 10:32:02.640 & -00:08:25.66 \\
2015A & Blind15A\_45 & 10:27:25.440 & +01:51:36.22 \\
2015A & Blind15A\_46 & 10:36:39.840 & +01:51:36.22 \\
2015A & Blind15A\_47 & 10:41:31.200 & +00:00:00.00 & Blind14A\_19 \\
2015A & Blind15A\_48 & 10:36:40.217 & -02:05:44.81 & Blind14A\_16 \\
2015A & Blind15A\_49 & 10:31:47.285 & -04:11:29.62 & Blind14A\_15 \\
2015A & Blind15A\_50 & 10:36:35.527 & -06:17:14.42 & Blind14A\_14

  \enddata
\label{tab:2015Afields}
\end{deluxetable}

\clearpage

\end{document}